\documentclass{optica-article}

\journal{opticajournal} % for journals or Optica Open

\articletype{Research Article}

\usepackage{lineno}
\usepackage{subfigure}
\usepackage{comment}
\usepackage{enumitem}
\usepackage{physics}
\usepackage[capitalise]{cleveref}
\usepackage[version=4]{mhchem}

\newcommand{\um}{{\textmu}m} % use enclosing parentheses {}. Otherwise it will eat up the following space.
\newcommand{\us}{{\textmu}s} % use enclosing parentheses {}. Otherwise it will eat up the following space.
\graphicspath{{figures/}}

\begin{document}

\title{Profile control of fibre-based micro-mirrors using adaptive laser shooting with \textit{in situ} imaging}

\author{Shaobo Gao \authormark{1,*}, Vishnu Kavungal \authormark{1}, Shuma Oya \authormark{1}, Daichi Okuno \authormark{1}, Ezra Kassa \authormark{1}, William J. Hughes \authormark{2}, Peter Horak \authormark{2}, and Hiroki Takahashi \authormark{1,+}}

\address{\authormark{1} Experimental Quantum Information Physics (EQuIP) Unit, Okinawa Institute of Science and Technology, 1919-1 Tancha, Onna-son, Okinawa, 904-0495, Japan\\
\authormark{2} Optoelectronics Research Centre, University of Southampton, Southampton SO17 1BJ, UK\\}

\email{\authormark{*}shaobo.gao@oist.jp}
\email{\authormark{+}hiroki.takahashi@oist.jp}
%% email address is required; see note below about the corresponding author designation

% use {asbstract*} to suppress the copyright line. Copyright information will be added in production

\begin{abstract*} 
Fibre Fabry-Perot cavities (FFPCs) are used in various studies in cavity quantum electrodynamics (CQED) and quantum technologies due to the cavity's small mode volume and compact integration with optical fibres. We develop a novel \ce{CO2} laser machining method that produces well-controlled surface profiles on the end facets of cleaved optical fibres. Using multiple shots in distinct spatial distribution patterns, our method employs a shooting algorithm that adaptively changes laser ablation parameters during the shooting to suppress deviations from the desired profile. This is made possible by \textit{in situ} imaging of the machined profile, its inspection and the usage of the information in the subsequent steps. Underlying this algorithm is a newly found laser ablation parameter, the pause between shots, which controls the accumulation of heat in between successive laser shots and as a result determines the area of impact made by an individual ablation sequence. We fabricate fibre-based micro-mirrors with radii of curvature ranging from 250~{\um} to 700~{\um} with an effective mirror diameter of 60~{\um} in either Gaussian or spherical profiles. Due to the self-correcting nature of our adaptive algorithm, we achieve a near 100\% success rate in the production of desired profiles with low ellipticity. After furnishing the laser machined fibre end facets with high reflectivity coating, FFPCs are formed to demonstrate a high finesse up to 150,000 at an optical wavelength of 854~nm. 
\end{abstract*}

%%%%%%%%%%%%%%%%%%%%%%%%%%  body  %%%%%%%%%%%%%%%%%%%%%%%%%%
\section{Introduction}
\label{sec:intro}

Fabry-Perot optical micro-cavities with high finesse are essential in modern quantum cavity electrodynamics (CQED). They have been utilised in numerous CQED experiments with atoms \cite{Volz2011,Gallego2018, Brekenfeld2020}, ions \cite{Steiner2013, Takahashi2017, Takahashi2020}, solid state emitters \cite{Albrecht2013, Kaupp2016, Herrmann2023} and optomechanical objects \cite{Kashkanova2017, Tenbrake2024, Rochau2021, Agrawal2024}. Various quantum applications have been also demonstrated with optical micro-cavities, such as quantum metrology and sensing \cite{ma2023, Wagner2018, Saavedra2022}, single-photon generation \cite{Riedel2017, Nisbet_2011}, quantum networks and quantum computing \cite{Niemietz2021, Fernandez-Gonzalvo2023, Ward_2022, Covey2023}. 
In these past achievements, micro-cavities using optical fibres known as fibre Fabry-Perot cavities (FFPCs) have been predominantly used. FFPCs carry ideal characteristics such as small mode volumes, compact footprints and intrinsic fibre coupling \cite{Pfeifer2022}. In FFPCs, end facets of optical fibre tips work as cavity mirrors.
To machine the end facet of an optical fibre into a concave mirror substrate, ablation with a \ce{CO2} laser is widely used \cite{Hunger2010}. Fused silica, of which optical fibres are made, has strong light absorption at the wavelength of the \ce{CO2} laser (9-10 {\um}). Therefore, a strong pulse of a \ce{CO2} laser can melt and evaporate the fused silica surface, creating a concave depression with surface roughness below 0.2 nm \cite{Hunger2012}, resulting in a high finesse cavity mirror after high reflectivity (HR) coating.

After the pioneering work of \cite{Hunger2010}, the \ce{CO2} laser machining of optical fibres has been studied and further developed by several groups to enable precise control of the machined profile \cite{Takahashi2014, Ott2016, Ruelle2019b} despite the complexity of the laser ablation mechanism \cite{Feit2010, Doualle2016}. It was understood that the symmetry of the profile is important to control the birefringence of the cavity \cite{Uphoff_2015}, and that the Gaussian profile imprinted by the Gaussian intensity distribution of the laser beam makes the cavity finesse sensitive to the cavity misalignment in comparison to a spherical profile \cite{Hughes2023}. Gaussian profiles, however, can still be favourable in some scenarios: they can form a stable cavity beyond the nominal stability limits defined for the case of the spherical mirrors with the same radii of curvature at the mirror centers (see \cref{sec:cavity_finesse}). Cavities with Gaussian mirror profiles can also increase coupling to an emitter through interference effects \cite{Podoliak_2017}. 

Although the symmetry issue can be solved by rotating the fibre while shooting with multiple laser shots \cite{Takahashi2014, Kay2024}, these methods only result in Gaussian surface profiles. The \textit{dot milling} technique introduced in \cite{Ott2016} achieves a more versatile control of the profile. In \cite{Ott2016}, it was shown that by engineering a spatial distribution pattern of the laser shots, one can achieve a spherical profile with good symmetry over a large area of the fibre end facet. Nevertheless, in \cite{Ott2016} and in the subsequent work \cite{Garcia2018}, the strategy for the dot milling was explained on an empirical basis, which is difficult to adapt in other settings, especially when one wants to tune the mirror shape between the Gaussian and the sphere. Besides, the authors did not explicitly provide a measure to counteract the potential instability of the laser machining caused by the nonlinearity and high sensitivity of the ablation process to experimental parameters. 
In this paper, we address these issues by first finding ablation parameters that are effective and easy to use, and elucidating their relations to the ablation result. Then we devise an adaptive laser shooting algorithm with \textit{in situ} imaging. The algorithm divides the machining process into several distinct steps, and between these steps, the machined profile is obtained from the imaging. The feedback from this imaging to the subsequent shooting steps enables us to automatically correct the asymmetry and deviations from the targeted profile caused by the imperfection and fluctuation of the experimental parameters.
We also demonstrate an easy switch between producing Gaussian and spherical mirror profiles with a simple off-line adjustment of the shooting steps.

Our fabrication technology will be useful for a wide range of applications based on Fabry-Perot resonators. To be specific, in this work we focus on cavities where the cavity length and the mirror radii of curvature are of the order of a few hundred micrometres, suitable e.g. for integration with ion traps \cite{kassa2025} and atomic ensembles \cite{Lucas2024}. To suppress the diffraction and the scattering losses at the mirror, a large mirror diameter $D$ with low surface roughness $\sigma$ are generally required \cite{Gao2023, Hughes2023}. Furthermore, the ellipticity, which leads to cavity birefringence \cite{Benedikter2015}, is also important. It can be favourable or not favourable to have a certain ellipticity depending on the application \cite{Kassa_2023, Barrett_2020}. In our paper, the profile ellipticity is characterized by the eccentricity defined as 
\begin{align}
r_e = \sqrt{1-R_{\text{minor}}/R_{\text{major}}},
\end{align}
where $R_{\text{major}}$ and $R_{\text{minor}}$ are the radii of curvature of the mirror profile along the major and minor axes, respectively. In this paper, we aim at a small $r_e$ and have achieved $r_e < 0.2$ with a reproducibility rate close to 100~\%. 
$r_e$ is related to the frequency splitting $\Delta\nu$ between two orthogonal polarisation modes \cite{Uphoff_2015} as
\begin{align}
    \frac{\Delta\nu}{\nu_{\text{FSR}}} = \frac{\lambda}{R_{\text{major}}}\qty(\frac{r_e}{2\pi})^2,
\end{align}  
where $\nu_{\text{FSR}}$ is the free spectral range (FSR) of the cavity and $\lambda$ is the wavelength of light. Therefore, $r_e < 0.2$ corresponds to $\Delta\nu$ being smaller than the resonance bandwidth of a cavity with a finesse up to 578,000 when $R_{\text{major}} = 500$ \textmu m and $\lambda = 854$ nm.

The structure of the paper is as follows: In \cref{sec:exp_setup} we first describe our experimental set-up. In \cref{sec:ablation_characterisation} we characterize the depth and width of the depression on a coreless fibre tip created by a single ablation sequence of the \ce{CO2} laser. Their relations to the pause between shots $\tau_{\text{PBS}}$ (explained later) and the number of shots $N$ are investigated. Our adaptive shooting algorithm is then introduced in \cref{sec:algorithm_results} based on the knowledge of the characterization in \cref{sec:ablation_characterisation}. Fibre micro-mirrors with Gaussian profiles produced with such an algorithm are used to form an FFPC. Its finesse as a function of the cavity length is reported in \cref{sec:cavity_finesse}. Preparation procedures for fibres in advance of the laser shooting and details of the data analysis for the cavity finesse are given in the supplemental document. 

\section{Experimental setup}
\label{sec:exp_setup}

\begin{figure}[h!]
\centering\includegraphics[width=8cm]{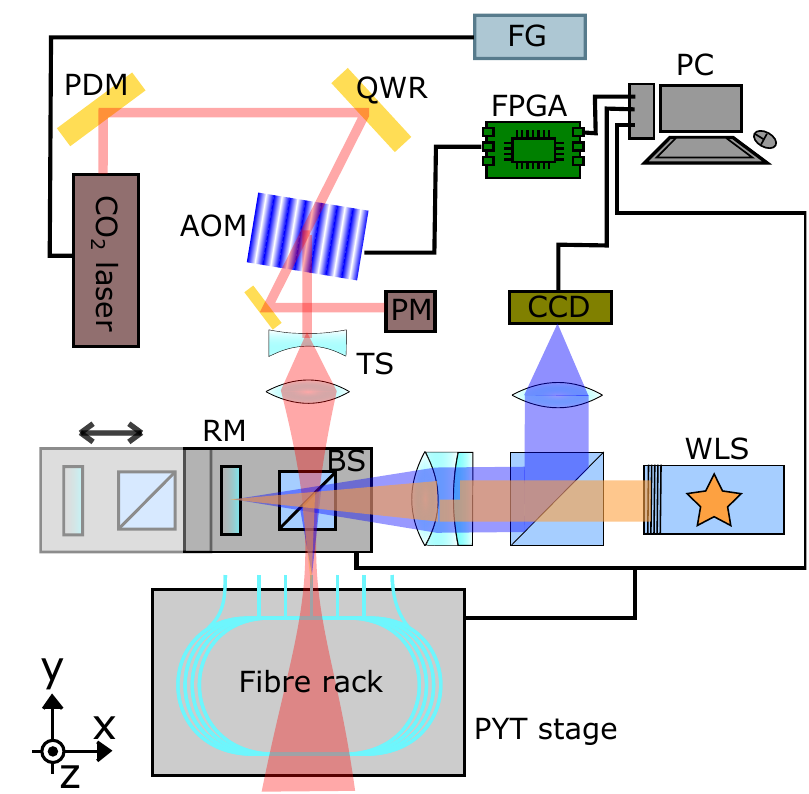}
\caption{A simplified schematic of the laser shooting setup. The red beam represents the \ce{CO2} laser. The orange and blue beams are the injected and back-reflected white light in the interferometer respectively. For details, see the main text.
AOM: acousto-optic modulator, BS: beam splitter, FPGA: field programmable gate arrays, CCD: charge coupled device camera, FG: function generator, PDM: polarisation dependent mirror, PM: power meter, PYT stages: pitch-yaw and translation stages, QWR: quarter-wave retarder, RM: reference mirror, TS: telescope lenses, WLS: white light source.}
\label{fig:setup}
\end{figure}

The experimental setup consists of four subsystems (see \cref{fig:setup}). The \ce{CO2} laser system, the positioning system, the imaging system, and the white light interferometer (WLI). The \ce{CO2} laser (SYNRAD v30) operates at 9.3~{\um} where the light absorption of silica glass is higher than at the more commonly used \ce{CO2} laser wavelength of 10.6~{\um}, resulting in a shorter penetration depth and less rounding at the edges of the fibre \cite{Kubista2017}. The laser is pulsed at 5~kHz. Since the heat transfer in a fibre is much slower than the pulsing period, the laser is effectively continuous with its power adjusted by the duty cycle of the pulses. In our case, the average laser power at the ablation target is 1 watt. In the laser system, the polarisation of the \ce{CO2} laser is cleaned and set to circular polarization by an absorptive polarization dependent mirror (PDM) and a reflective quarter-wave retarder (QWR), before it is transmitted through an acousto-optic modulator (AOM: Brimrose GEM-40-2-10.6). The laser is circularly polarised to reduce the ellipticity of the single-shot ablation profile \cite{Uphoff_2015}. The AOM diffracts the \ce{CO2} laser beam to the first order with an efficiency of about 50~\%. This first-order diffraction is subsequently focused by a ZnSe telescope to a waist of less than 25~{\um} at the ablation target. By switching on and off the first-order diffraction of the AOM with a transient time in microseconds, a "shot" of the \ce{CO2} laser is sent onto the target. The duration of the shot is on the order of tens of milliseconds, comparable with the time of the heat transfer dynamics in a fibre. The power of the laser is monitored with the zeroth-order diffraction of the AOM output by a thermal power meter.

The positioning system is a stack of three translational stages (PI L-511, Attocube ECS5050x, PI L-306) for the movements along the $x$, $y$ and $z$ axes respectively and a pitch-yaw rotation stage (Thorlabs PY004Z8). 
On top of the stack the ablation target is mounted. In the case of shooting fibres a custom-made fibre rack is mounted to host up to eleven fibres (see the supplemental document). The pitch-yaw rotation stage is used to align the angle of the surface of the ablation target at the beginning of the machining process with respect to the reference mirror (RM) in the WLI which is in turn aligned perpendicular to the incoming \ce{CO2} laser beam in advance. During the machining process, the ablation target is translated in the $xz$ plane to allow for shooting at different transverse positions. The resolution and repeatability of the translation stage (PI L-511) are quoted to be 50 and 100~nm, respectively, which are small enough for our purpose. 
%The translation along the $y$-direction is used to sample interferograms at different longitudinal positions in the WLI.

The imaging system comprises a long working distance infinity-corrected objective (Mitutoyo M Plan Apo 10X) and a tube lens (Thorlabs-TTL200-A). White light is injected through the imaging system and the beam splitter (BS) to illuminate the ablation target as well as the RM. The WLI is a Michelson-type interferometer similar to the one implemented in \cite{Takahashi2014}, with the surface of the ablation target working as one of the end mirrors and the RM working as the other. The combined reflection is imaged at the CCD camera to produce an interferogram of the machined surface referenced to the flat surface of the RM. The RM and BS are mounted together on a translation stage (PI VT80). Using this translation stage, the BS moves away from the laser beam path during ablation to avoid distortion to the laser beam, and returns to the original position when profile reconstruction is performed. In this way, one can swiftly switch between the laser shooting and imaging/WLI configurations without moving a fibre, which enables us to develop an adaptive shooting algorithm that utilises feedback from imaging.
The profile reconstruction is carried out from six interferograms with the ablated target translated along the imaging direction ($y$-axis) at six different positions incremented by $\lambda/8$, where $\lambda$ is the central wavelength of the white light \cite{Schmit1995, Xia2016}. The exact algorithm used to reconstruct the surface profile is described in reference \cite{Zhao2019}, which performs fast and robustly against imaging noise. 

Before ablation, to calibrate the system, the $\text{CO}_2$ laser beam size at the target position was measured with a "knife-edge" measurement, where a blade moves across the beam transversely and clips off the transmitting power. To determine the beam size, the transmission versus the blade's transverse position is fitted by an error function. To align the imaging system to the laser position, we first apply a single ablation attempt on a glass plate and then align the imaging system to the crater produced on the glass plate. The alignment is then fine-tuned by ablating on a fibre. A Matlab GUI program acquires images from the CCD camera, and controls the stages and an FPGA (Opal Kelly XEM3001v2) that produces pulses to switch on and off the AOM with well-defined timings. All processes can be automated in a preprogrammed sequence except for the fibre preparation and loading, where a vacuum-compatible single-mode fibre is spliced with 20~{\um} long coreless fibre, to avoid the effect of the discontinuity between the fibre core and cladding \cite{Takahashi2014}. See the supplemental document for more details. 

\section{Ablation Characterization}
\label{sec:ablation_characterisation}

\begin{figure}[ht!]
\subfigure[]
    {
    \includegraphics[width=0.45\textwidth]{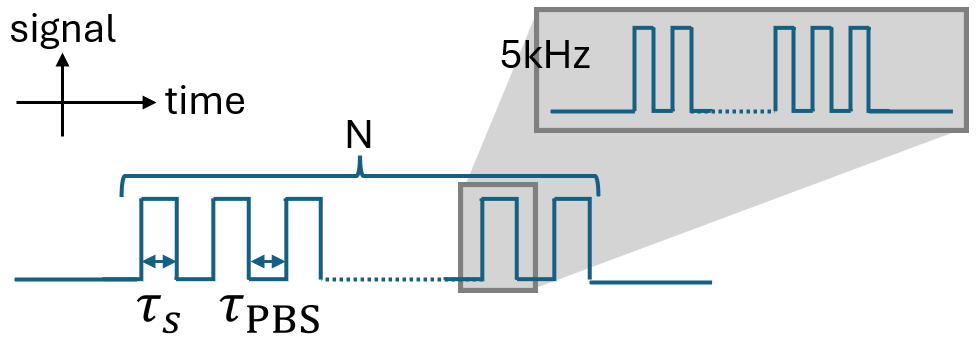}
    \label{fig:pulses}
    }
    \subfigure[]
    {
    \includegraphics[width=0.4\textwidth]{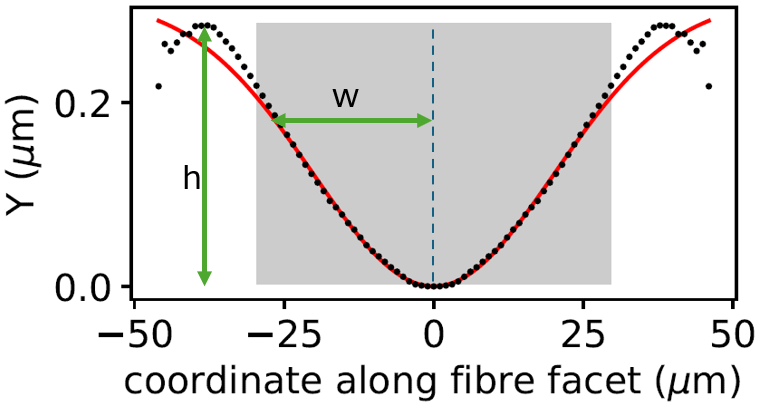}
    \label{fig: Gaussian fit}
    }
    \\
    \subfigure[]
    {
    \includegraphics[width=0.45\textwidth]{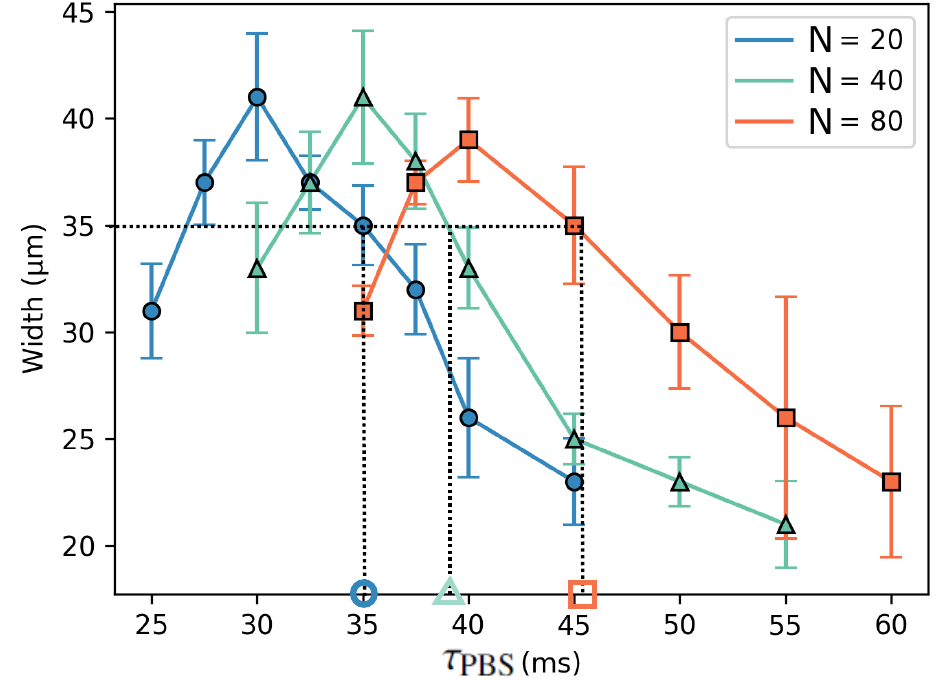}
    \label{fig:ablation char_a}
    }
    \subfigure[]
    {
    \includegraphics[width=0.45\textwidth]{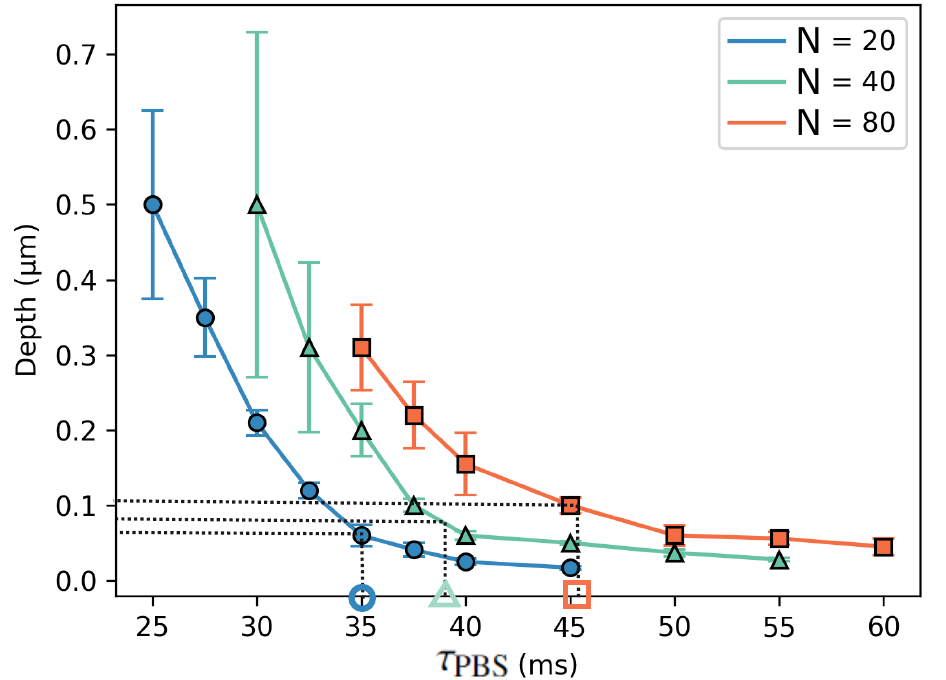}
    \label{fig:ablation char_b}
    }

\caption{\textbf{(a)} A single ablation sequence: the sequence is a train of $N$ shots with a shot duration $\tau_{s}$ (around 10~ms) and a pause between shots $\tau_{\text{PBS}}$. Each shot consists of fast pulses at 5~kHz. \textbf{(b)} Cross section of an ablation depression (black dots) fitted to a 2-D Gaussian function (red line). $h$ is the depression depth, measured by the distance between the profile's peak and valley. $w$ is the width of the depression obtained from the Gaussian fitting (with an amplitude of $h$). Grey shade indicates the region of interest for the fit (30 \textmu m in radius). \textbf{(c)} Ablation depression width $w$ as a function of the pause between shots $\tau_{\text{PBS}}$ and for different shot numbers $N=20,40,80$. \textbf{(d)} Ablation depression depth $h$ as a function of $\tau_{\text{PBS}}$ and for $N=20,40,80$. In (c) and (d) the error bars are the statistical standard deviations obtained from measuring different samples. Settings to mill a 35~{\um} feature are indicated by horizontal and vertical dotted lines.}   
\end{figure}

We apply multiple laser shots on the tips of the fibres for gradual and controllable ablation. At each shooting position, we irradiate a fibre with a sequence of $N$ laser shots in which each shot lasts a duration $\tau_s$ and consecutive shots are separated by a pause $\tau_{\text{PBS}}$. We call each of these sequences a \textit{single ablation sequence} (see \cref{fig:pulses}). The sequence is produced by sending voltages from the FPGA to the AOM in the corresponding temporal profile. As mentioned previously, each shot comprises pulses at a 5~kHz repetition, the duty cycle of which determines the ablating power. We ignore this fast pulsing as a variable in the rest of the paper. On the other hand, a "shot" has a much slower time scale of about 10~ms. Therefore its duration ($=\tau_s$) and mutual interval ($=\tau_{\text{PBS}}$) play crucial roles during the ablation process. It is important to understand how each ablation sequence modifies the fibre tip surface. 

A single ablation sequence creates a depression that can be approximated by a Gaussian function, as shown in \cref{fig: Gaussian fit}. We characterize the depression with two values: the height difference $h$ between the top and bottom of the depression and the width of the depression $w$. Both are characterised by a fit using a 2D Gaussian function $h\cdot[1-\text{exp}(-r^2/w^2)]$, where $r$ is the radial distance measured from the centre of the depression (see \cref{fig: Gaussian fit}).
As is the case in all the Gaussian fittings in this paper, the Gaussian amplitude $h$ is obtained from the profile's peak and valley, and then the waist $w$ is fitted with the fixed amplitude $h$.
The radius of curvature (RoC) of the profile created with a single shot is carefully discussed in reference~\cite{Ruelle2019}. In laser machining using multiple shots like ours, however, the RoC of every single shot plays a less direct role, so we exclude the discussion of this value. 

The experimental parameters in our control are as follows: laser power $P_l$, laser beam size $D_l$, number of shots $N$, shot duration $\tau_s$, and pause between shots $\tau_{\text{PBS}}$. We start with three general constraints and observations of the parameters:
\begin{enumerate}
\item	It is experimentally challenging to change the laser beam size $D_l$ at the shooting position dynamically in the time scale of milliseconds.

\item The laser beam diameter $D_l$ and the fibre diameter set the upper limit for the depression width radius $w$.

\item	For the parameter region where the ablation occurs, the impact of $P_l$ and that of $\tau_s$ are similar. This is true when $\tau_s$ is similar to or smaller than the heat dissipation time, therefore in such a condition, the effects of changing $\tau_s$ can be substituted by changing $P_l$, and vice versa.  In this regime, the heat transfer does not play a dominant role within a single shot, and hence what matters during the ablation is the total energy that is absorbed by the fibre material.
\end{enumerate}

With these considerations, we can largely simplify the parameter space. Based on the points 1 and 2, we fix the laser beam diameter $D_l$ to be comparable to the expected effective fibre diameter at the completion of the laser machining, which is around 80 {\um} (our fibre diameter is 125 {\um}, its tip will shrink and be rounded after laser machining). Based on point 3, we fix the laser power (to be around 1W) and instead adjust only the shot duration $\tau_s$ to vary the energy injected by a single shot. Changing the laser power is less favourable than changing $\tau_s$ because even though the former can be realized either by changing the duty cycle of the laser or changing the AOM drive amplitude, they would cause instability of the laser or suffer from the difficulty in control due to the nonlinear dependence of the AOM's diffraction efficiency on the AOM drive amplitude, respectively. We adjust $\tau_s$ so that each shot has just enough energy to evaporate a small amount of the material. In our experiment, we found that around 9.6 ms works well. However, this number is highly dependent on the exact \ce{CO2} laser power and laser beam diameter. It is easy to find an appropriate $\tau_s$ experimentally once the laser power $P_l$ and the laser beam diameter $D_l$ are fixed. One note is that since the ablation process is highly nonlinear, a small change of $\tau_s$ could lead to either complete melting of the fibre tip or no ablation at all. Then we are left with just two parameters $N$ and $\tau_{\text{PBS}}$ that can be adjusted easily, precisely, and fast with the AOM.

\cref{fig:ablation char_a} and \cref{fig:ablation char_b} show the relationship between the ablation results ($h$, $w$) and the ablation parameters ($N$, $\tau_{\text{PBS}}$). The exact impacts also vary with the shot time $\tau_s$, but the trends are the same as long as the individual impact of the shot is small enough. We observe one important and useful feature from the results: The depression width $w$ increases for small values of $\tau_{\text{PBS}}$, reaches a maximum at a critical time $T_{\text{crt}}$, and then decreases again.
The qualitative physical explanations are as follows: During a shot, heat accumulates until the fused silica reaches the evaporation temperature. The central region of the beam will accumulate most of the heat because the laser has a Gaussian intensity profile. As mentioned earlier, $\tau_s$ is adjusted so that once the centre region starts to evaporate, the shot terminates, so the impact of the ablation is localized at the centre of the laser spot. During the pause between shots, the heat dissipates away on tens of milliseconds time scale. If the pause is long enough, the temperature of the fibre resets close to the room temperature and the impact of consecutive multiple shots is the simple additive effect of individual shots, which does not change the width of depression $w$. As a result, the width $w$ remains small. However, if the pause is not long enough to reset the temperature but long enough to equalize the temperature across the fibre tip surface through heat transfer, the following shot will take less time to heat the fibre tip up to the evaporation temperature than using an isolated single shot. This leads to evaporation in a larger area on the fibre tip, resulting in a larger width $w$. This explains the width $w$ decreasing with $\tau_{\text{PBS}}$ when $\tau_{\text{PBS}}>T_{\text{crt}}$. When $\tau_{\text{PBS}}<T_{\text{crt}}$, the evaporating area becomes comparable with the fibre diameter, the fibre edge will curl up and the fibre shrinks due to surface tension, effectively decreasing the radius of the depression as seen in \cref{fig:ablation char_a}. We call this an "edge-curling" effect, which causes the previously mentioned phenomenon that the depression width radius $w$ decreases with $\tau_{\text{PBS}}$ when $\tau_{\text{PBS}}>T_{\text{crt}}$ but increases when $\tau_{\text{PBS}}<T_{\text{crt}}$.

With this feature, we can fix all the other parameters ($P_l$, $D_l$ and $\tau_s$), and choose the depth and width of the depression by adjusting $\tau_{\text{PBS}}$ and $N$ correspondingly. For example, if we want to have a local depression with a width of 35~{\um} and depth of 100~nm, firstly from \cref{fig:ablation char_a}, there are three ($N$, $\tau_{\text{PBS}}$) configurations to give the width  35~{\um}, indicated by circle, triangle and square of three different colours. These configurations correspond to three different depression depths shown in \cref{fig:ablation char_b}. One can select the one closest to 100 nm, in this case, $N = 80$. It is also easy to interpolate to a more accurate value of $N = 75$. Our shooting algorithm is based on the tuning of these two parameters.

To the best of our knowledge, our work has revealed the role played by $\tau_{\text{PBS}}$ in the \ce{CO2} laser ablation quantitatively for the first time. $\tau_{\text{PBS}}$ provides an important degree of freedom to adjust the impact made by a single ablation sequence, which was not explored previously. Without $\tau_{\text{PBS}}$, one is left with either $P_l$ or $\tau_s$ as a variable parameter if $D_l$ is fixed. As already noted, the ablation profile is highly sensitive and non-linear to both $P_l$ and $\tau_s$ while the resolution of $\tau_s$ is also limited by the pulsing period of the \ce{CO2} laser (= 200~{\us}). On the other hand, the change incurred by the variation of $\tau_{\text{PBS}}$ is more moderate. Its time scale is on the order of milliseconds due to the underlying thermal process, and it is not limited by the pulsing period as it controls the absence of the laser. These properties make it relatively easy to control $\tau_{\text{PBS}}$ together with $N$ to attain a desired local impact by each ablation sequence.      

\section{Adaptive shooting algorithm and results}
\label{sec:algorithm_results}

The dot milling approach, in which the fibre is moved in the lateral plane perpendicular to the laser beam and shot at different positions, allows one to design a machined profile as demonstrated in \cite{Ott2016}. 
However, since the laser ablation process is highly non-linear, small fluctuations in the experiment, e.g. laser power, laser pointing, fibre position change, and any unknown reasons that disturb the heat flow of the fibre, would change the results. So, any fixed shooting pattern or ablation parameters in the dot milling cannot guarantee the success of every attempt. 

The capability to dynamically adjust the ablation width and depth, while simultaneously acquiring the surface profile during machining, allows us to develop a robust laser shooting algorithm that compensates for such experimental errors. The algorithm divides the total laser machining process into several distinct steps. With \textit{in situ} imaging, the ablation result of each step is automatically captured without moving the fibre. The ablation result of a prior step determines the shooting pattern in the next step, such that errors in the prior step are compensated. Hence, the algorithm proceeds adaptively. Before each adaptive compensation, the reconstructed surface profile is processed with Gaussian filtering with a standard deviation of $\approx$ 3.5 ~{\um} to eliminate high spatial frequency fluctuations of the surface. We believe such fluctuations are due to the noise in the imaging. In the later section, this assumption is verified by forming a high-finesse cavity with fibre mirrors fabricated using this algorithm. If the fluctuations were on the mirror surfaces, they would have resulted in a significantly lower cavity finesse. This filtering helps increase the accuracy of the curvature and centre position in the fitting.

\begin{figure}[h!]
    \centering
    \includegraphics[width=13cm]{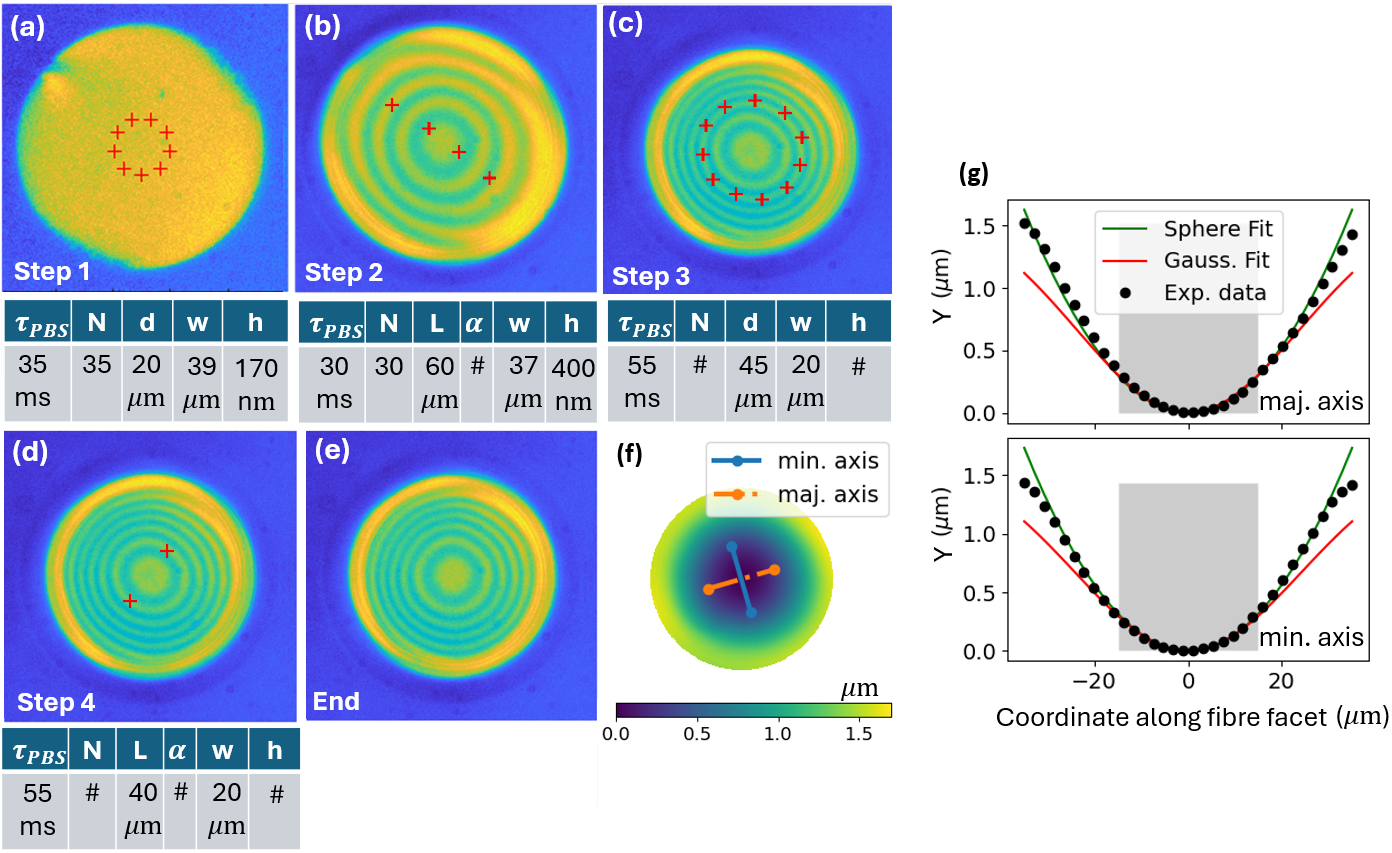}
    \caption{Exemplary results of the laser shooting algorithm: \textbf{(a)-(e)} White light interferograms of the fibre end facet prior to distinct shooting steps are displayed with red markers indicating the positions to be shot. Note that the interferograms reveal the outcomes of the prior shooting steps. For example, the interferogram in (b) corresponds to the result of the shooting step 1 indicated by the circular pattern in (a). The interferogram in (a) represents the flat fibre end facet after cleaving. The tables below the images list the shooting parameters used in each step. The parameter $d$ in (a) (Step 1) and (c) (Step 3) is the diameter of the circular shooting pattern; $L$ in (b) (Step 2) and (d) (Step 4) is the total length of the linear pattern; $\alpha$ in (b) (Step 2) and (d) (Step 4) is the angle between the linear pattern and horizontal axis; "$\#$" means that the value of the corresponding parameter is automatically determined by the algorithm described in the main text. $w$ and $h$ are the ablation depth and height introduced in Section~\ref{sec:ablation_characterisation}. \textbf{(f)} Heat map of the reconstructed surface profile after the completion of all the shooting steps, obtained from the interferogram in (e). The major and minor axes (maj. axis and min. axis) are extracted by fitting an ellipsoid to the profile. \textbf{(g)} Gaussian and sphere function fittings are applied along the major and minor axes of the reconstructed profile in (f). The grey zones indicate the fitting areas in the algorithm, which, compared to \cref{fig: Gaussian fit} is smaller because the algorithm focuses more on the quality of the central region.}
\label{fig:algorithm}
\end{figure}

The algorithm has four main steps: global ablation, global correction, spherical correction (optional), and central correction. \cref{fig:algorithm} shows the result at each step in an exemplary laser machining run. The idea behind the distinction of these steps is as follows: In the first global ablation step, the fibre is ablated while setting the ablation width $w$ relatively large. This step efficiently creates an overall structure across the entire end facet. The subsequent three steps make corrections to the imperfections of the machined profile in different aspects. The second ``global correction'' step corrects the overall elliptic asymmetry of the profile. The prior global ablation step normally ends with a large profile asymmetry. However, since in the global ablation $w$ is set comparable to the effective fibre size, the profile is free from small-scale local variations. This allows us to correct the asymmetry with another ablation step with a large $w$. The third ``spherical correction'' modifies the profile from a Gaussian-like shape to close to a spherical shape by locally ``digging'' the fibre surface. In the final ``central correction'' step, the ellipticity left in the central region is removed. Note that the ablation width $w$ is set relatively small in the last two steps than in the first two steps to localise the ablation impact (see \cref{fig:algorithm}). A Gaussian shape can be achieved by simply omitting the spherical correction step. In each step, shooting parameters that are used to compensate random errors caused by the environmental fluctuations are labelled as “\#” in the tables of Fig.3, which indicate values are to be determined in situ. The details of each step are elaborated in the following.

\begin{enumerate}[label={Step \arabic*.}]
    \item Global ablation: The fibre tip is shot in a circular pattern. The $\tau_{\text{PBS}}$ is chosen to maximise $w$, and thus to have a global profile change. The circular pattern increases the profile's radius of curvature compared with a single ablation at the centre. The circle diameter is chosen such that the distance from the shooting spots to the edge of the fibre is approximately equal to the ablation radius $w$, this maximises the ablation region with negligible "edge-curling effect", which in our case is 20~{\um}. The number of ablation positions, and the number of shots per position $N$ are chosen such that the ablation depth is roughly half the final depth. We decided the number of the shooting positions to be 10 and $N$ to be 35 so that the depth at the centre is around 800 nm. An example of the ablation pattern and its outcome are shown in \cref{fig:algorithm} (a) and (b), respectively.
    \item Global correction: The correction is in a linear pattern along the major axis of the elliptic profile. The major axis is calculated \textit{in situ} from the reconstructed profile. $\tau_{\text{PBS}}$ is slightly shorter than the ``global ablation'' step to have an edge-curling effect as mentioned in \cref{sec:ablation_characterisation}. The end positions of the linear pattern should be chosen such that the ablation width $w$ covers the edge of the fibre. In the case of \cref{fig:algorithm}, $\tau_{\text{PBS}}$ of 30~ms corresponds to a width $w$ of 35~{\um}. Thus, the end positions of the linear pattern are located at 35~{\um} from the fibre edge, resulting in the length of the pattern being around 60~{\um}. This edge-curling effect will decrease the radius of curvature along the profile's major axis and thus correct the asymmetry, as shown in \cref{fig:algorithm} (c). The shooting number $N$ is chosen so that the depression depth matches the target result.
    \item Spherical correction: The profile is naturally close to a Gaussian shape after the previous steps. To amend this to a spherical shape, we apply a circular pattern with a long $\tau_{\text{PBS}}$ that offers a small ablation radius $w$. We choose a relatively small $N=30$ with $\tau_{\text{PBS}} = 55$~ms for a minimum and controllable impact. Once the circle is completed, the profile is inspected. Then this process is iterated until the error from a spherical shape becomes acceptable. The diameter of the circular pattern $d$ can be optimised as follows: the expected profiles after this correction step are precalculated while sweeping $d$ for a given set of $w$ and $h$. The optimal $d$ is chosen so that it results in the least deviation from the targeted sphere shape. In practice, however, such a calculation empirically always settles down to a similar value of $d$. Hence, we fixed it to be 45~{\um} as in \cref{fig:algorithm} to save the calculation time. The number of the shooting positions is chosen such that the distance between adjacent shooting positions is roughly equal to the ablation width $w$, resulting in 10 positions in \cref{fig:algorithm}.

    \item Central correction: The central region is the most important region where the cavity mode is located. It is crucial to guarantee the profile symmetry at the centre. So in the final step, we apply two local ablations along the minor axis close to the centre to correct the asymmetry at the centre, as shown in \cref{fig:algorithm} (d). The $\tau_{\text{PBS}}$ is chosen to be the same as in the prior spherical correction step to have localised impacts. The distance between the shooting positions and the fibre centre is approximately equal to the ablation width $w$, which is 20~{\um}. This process is iterated until the ellipticity reaches the goal, in our case, $r_e < 0.2$.
\end{enumerate}

\begin{figure*}[h!]
\centering
    \subfigure[]
    {
    \includegraphics[width=0.302\linewidth]{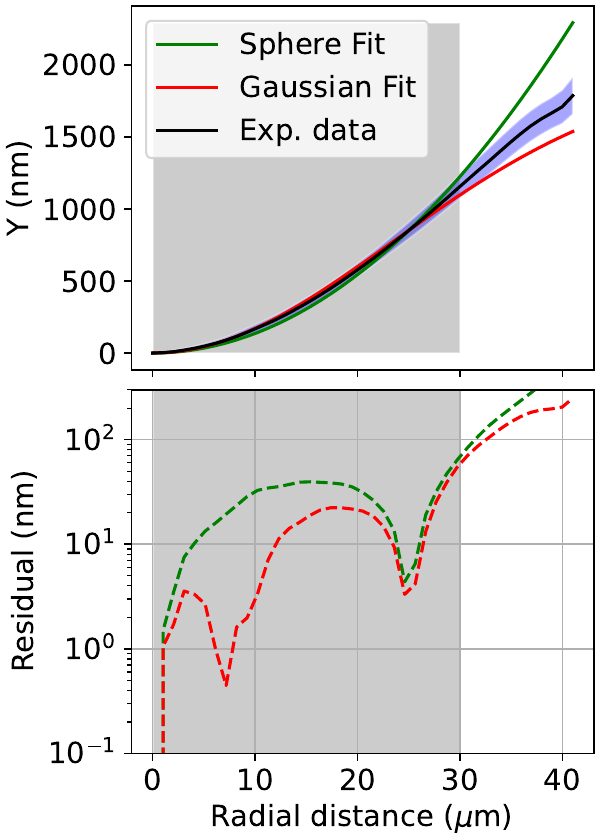}
    \label{fig:Gaussian profile}
    }
    \subfigure[]
    {
    \includegraphics[width=0.243\linewidth]{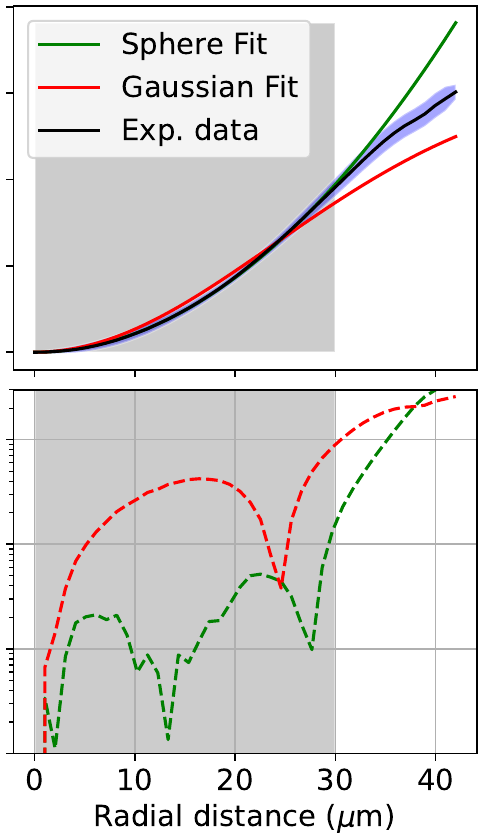}
    \label{fig:sphere profile}
    }
    \subfigure[]
    {
    \includegraphics[width=0.297\linewidth]{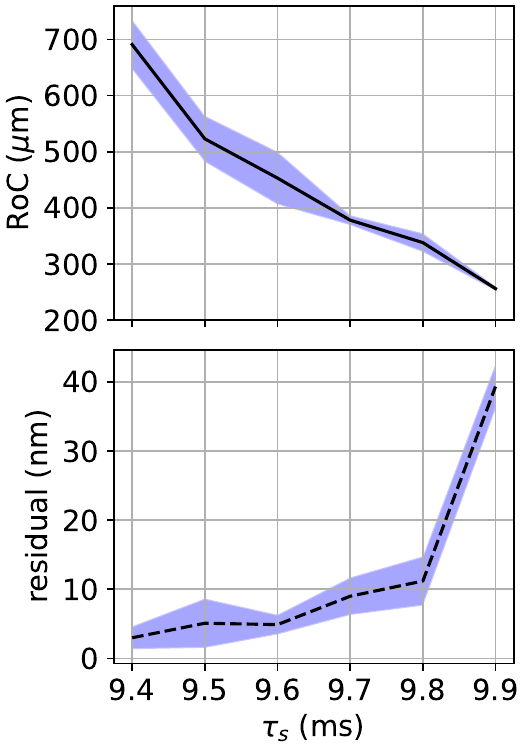}
    \label{fig:RoC tune}
    }

\caption{\textbf{(a)} (Top) The spherical and Gaussian function fitting to a profile produced with the adaptive algorithm without the spherical correction step (Step 3 in \cref{fig:algorithm}). The 1D experimental data (black line) is obtained in such a way that the 2D profile is rotationally averaged around the central axis perpendicular to the end facet. The blue-shaded error bar is the standard deviation of the data points at a constant radial distance from the centre. The grey shade is the region used for the data fitting (30 \textmu m). (Bottom) The red and green dashed lines are the residuals of the spherical and Gaussian fittings from the 1D-averaged experimental data (black line), respectively. \textbf{(b)} The same fittings as in (a) are applied to a profile produced while incorporating the spherical correction step. For a fair comparison the same fibre sample as (a) was used. See the main text for the details. \textbf{(c)} (Top) The radii of curvature of spherical profiles as a function of the shot duration $\tau_s$. The shade indicates the uncertainty of the RoC. (Bottom) The residual from the spherical fit is shown in the dashed line. The residual is calculated within the fitting region (30 \textmu m in radius). The blue-shade error bars for both top and bottom plots are the standard deviations taken from different fibre samples.}
\end{figure*}

\cref{fig:Gaussian profile} and \cref{fig:sphere profile} demonstrate fitting of the Gaussian and spherical functions to laser-machined fibre surface profiles. The profile used in \cref{fig:Gaussian profile} is produced without the spherical correction step in the algorithm. 
As a result, the Gaussian function fitting gives a smaller residual of the fit from the experimental data than the spherical one.
This relationship is reversed in \cref{fig:sphere profile} where the profile is produced while incorporating the spherical correction step. 
For a fair comparison, the same fibre sample was used in \cref{fig:Gaussian profile} and \cref{fig:sphere profile}: first the fibre was laser-machined without the spherical correction step to produce the data in \cref{fig:Gaussian profile}. Then the fibre was shot again with the spherical correction followed by the central correction, and the data in \cref{fig:sphere profile} was obtained.         
With the spherical correction step, the profile can be approximated to be a sphere up to a radius of 30~{\um} as seen in \cref{fig:sphere profile}, with deviation from the spherical fit to be $<10$~nm at any point within the region of interest, comparable with the state of the art in reference \cite{Ott2016, Kay2024}. By comparing \cref{fig:Gaussian profile} and \cref{fig:sphere profile}, we can conclude that the spherical correction significantly improves the effective sphere mirror diameter. 

Note that our current shooting algorithm does not include adaptive correction for the RoC of the profile. This is, in principle, possible by introducing additional correction steps in the algorithm. However, doing so would require a significant amount of additional laser shots which makes the fibre end facet further shrunk to a prohibitively small diameter. For this reason, currently we prioritise the corrections for the profile symmetry and sphericity in the algorithm, and the development of an algorithm that efficiently corrects the RoC in addition is left for the future work.       
Nonetheless the shot duration $\tau_s$ can be adjusted to change the RoC of the profile. As previously noted, the ablation result is sensitive to $\tau_s$. By changing $\tau_s$ from 9.4 ms to 9.9 ms, while keeping all the other parameters the same, the RoC can be changed from 250 {\um} to 700 {\um}. \cref{fig:RoC tune} shows the dependence of the RoC on $\tau_s$. Note that only results for spherical profiles are presented, as the results for Gaussian profiles are similar.

We can achieve an ellipticity <0.2 with almost 100 \% success rate if the fibre is prepared in a good condition, i.e. being flat and relatively defect-free after fibre cleaving. For example, we have had a consecutive run of fabrication where 75 fibres were successfully laser-machined out of 77 attempts. The two failures in this run were ascribed to sudden changes of the laser power during the machining. In total, so far we have laser-machined several hundred fibre end facets with various radii of curvature. One thing to note is that even though the success rate is very high, it is still challenging to control the radii of curvature precisely, as the uncertainties are shown as the error bars in \cref{fig:RoC tune}. We suspect that this is mainly due to laser power fluctuations in the current system. We believe it will be necessary to implement active laser power stabilisation to achieve good control over both profile eccentricity and RoC in the same process. 

\section{Cavity finesse measurement}
\label{sec:cavity_finesse}

\begin{figure}[h!]
\centering
\subfigure[]{
\includegraphics[width=5.5cm]{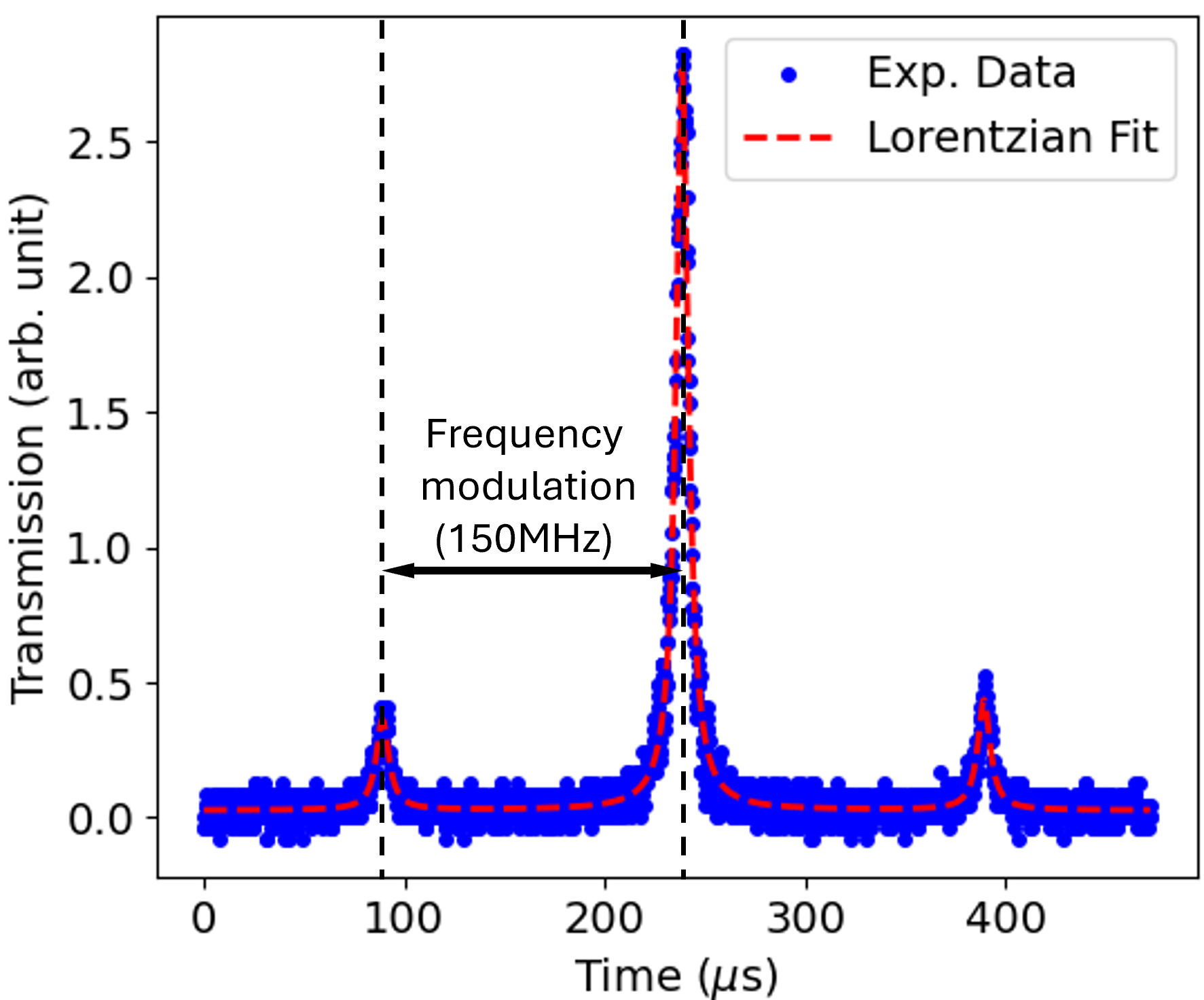}
\label{fig:cavity_spectrum}
}
\subfigure[]{
\includegraphics[width=7cm]{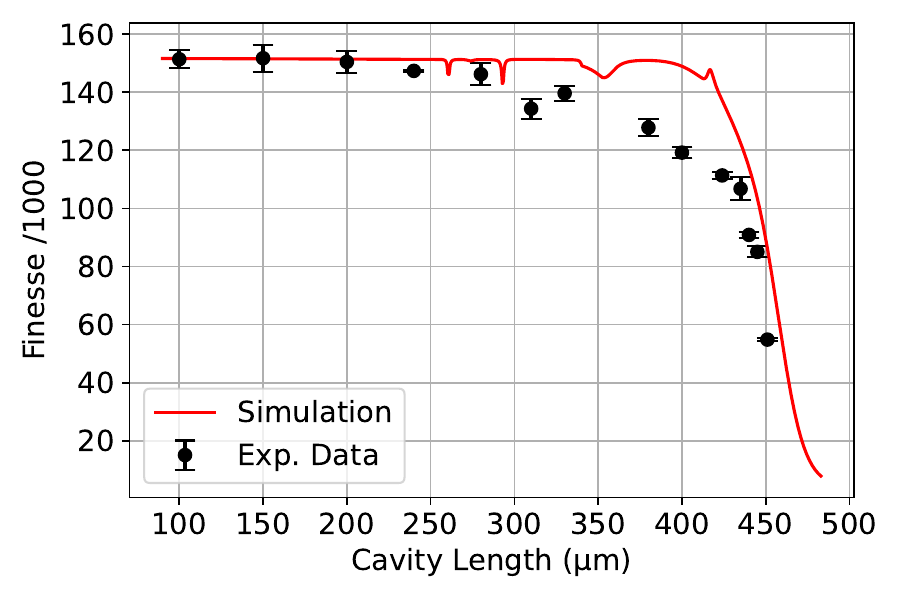}
\label{fig:finesse_vs_length}
}
\caption{\textbf{(a)} An example of the cavity spectrum when scanning the cavity length. The x-axis is the real-time register from the oscilloscope. The y-axis is the voltage output from a home-made photodiode detector. The sidebands of the spectrum originate from the laser frequency modulation with an electro-optical modulator (EOM). \textbf{(b)} Cavity finesse of an FFPC as a function of the cavity length. See the main text for the details of the mirrors. The error bars represent the standard deviations of six measurements in the same setup performed at different times. The red solid line is the simulation; see the main text and supplemental document for details.}
\label{fig:finesse}
\end{figure}

A set of laser-machined fibres were HR coated and subsequently tested by forming cavities. The coating was specified for the best possible reflectivity at a wavelength of 854~nm (LASEROPTIK GmbH), which is the wavelength of one of the atomic transitions in calcium ions. \cref{fig:finesse_vs_length} demonstrates the measurement of the finesse of one of such cavities as a function of the cavity length. We used Gaussian shaped mirrors in this measurement. The radii of curvature of the two fibre mirrors used in this measurement are 594~{\um} and 419~{\um}. The corresponding unstable region in the case of spherical mirrors with the same RoC is between the cavity lengths of 419~{\um} and 594~{\um}. One of the fibres was mounted on a ring piezo actuator (Piezomechanik HPSt150) for cavity-length scanning. An 854 nm continuous-wave laser was coupled into the cavity through the scanning fibre, and the cavity transmission of the fundamental cavity mode was measured from the output of the other fibre while scanning the cavity length. 
The laser was frequency-modulated with an electro-optic modulator (EOM), creating fixed sidebands at 50-200 MHz. The cavity resonance linewidth was calculated from the transmission spectrum with the sidebands as a frequency reference (see \cref{fig:cavity_spectrum}). The FSR of the cavity was calculated from the cavity length measured with an optical microscope. The cavity system was placed inside a home-made clean hood to minimise contamination. 

As shown in \cref{fig:finesse_vs_length}, the FFPC could maintain a high finesse above 100,000 up to a cavity length of 430 \textmu m, beyond the stability limit for the spherical mirrors with the same radii of curvature. This is the advantage when using Gaussian mirrors mentioned in \cref{sec:intro}. It can be understood as follows: the effective RoC sampled by the cavity mode of a finite size is larger with a Gaussian mirror than with the corresponding spherical mirror with the same central RoC. The effective RoC becomes even larger with the cavity length since the mode field size becomes larger on the Gaussian mirrors with the cavity length \cite{Benedikter2015}. As a result, the stability limit can extend beyond the one given by the radii of curvature measured at the mirror centers.
The finesse is well predicted by fitting the surfaces with Gaussian Process Regression and calculating the cavity modes using a mode-mixing approach, see supplemental document. We did not observe any transmission line splitting caused by birefringence, attesting to a good axial symmetry of the mirror profile. At short cavity length, a finesse of 150,000 is observed, corresponding to a round-trip loss of 40 ppm. The combined loss for scattering and absorption of the HR coating is 5-10 ppm, and the mirror transmission is 5 ppm according to the specifications from the company. This leaves only 5-10 ppm of extra scattering and/or clipping loss per mirror. This is consistent with the expected scattering loss ($\equiv S$) due to the fibre surface roughness of $\sigma \approx 0.2$~nm: from the well-known formula of $S\approx \qty(4\pi\sigma/\lambda)^2$ \cite{Bennett1992} , $S$ is calculated to be 8-9~ppm. 

\section{Conclusion}
\label{sec:conclusion}

Fabricating fibre-based micro-mirrors at various radii of curvature up to the millimetre scale is important for many quantum applications. We use a $\text{CO}_2$ laser to ablate a fibre end facet at multiple positions to produce such fibre mirrors. 
We identified two ablation parameters, the pause between shots $\tau_{\text{PBS}}$ and shot number $N$, amongst other parameters, to control the impact of a single ablation sequence. 
In particular, $\tau_{\text{PBS}}$ controls the ablation width, which provides an important degree of freedom that enables switching between global ablation and localised correction in the multiple-shot shooting algorithm. 
The algorithm is \textit{in situ}, meaning the profile is inspected during the laser-machining, and adaptive, meaning the surface profile after each step is used to determine the shooting parameters in the next step. In this way, the algorithm can accommodate erroneous fluctuations during the machining. 
We have machined more than 100 fibres with a nearly 100\% success rate of achieving low ellipticity ($r_e<0.2$). We can control the radii of curvature between 250~{\um} and 700~{\um} with modest tuning of the shot duration.
Furthermore, we can fabricate either spherical or Gaussian shape mirrors with a simple adjustment of the shooting steps. 
We have verified the quality of the mirrors by forming an optical cavity that shows a cavity finesses of more than 150,000, without resonance transmission splitting and finesse above 100,000 up to a cavity length of 430 {\um}. The mirror quality is comparable with the state of the art \cite{Ott2016,Kay2024} in terms of both surface roughness and profile accuracy. A robust, versatile and easy production of such fibre mirrors facilitates various experimental studies as well as technological applications that involve light-matter interactions, such as quantum technology with neutral atoms \cite{Lucas2024}, ions \cite{kassa2025}, and quantum dots \cite{Yang2024}, or even chip-based optical environmental sensing \cite{Javernik2019}. Our fabrication technologies can also be applied to other mirror substrates than optical fibres, such as glass templates \cite{Doherty2023}. If combined with an improved fibre pre-shooting preparation procedure (Supplementary 3), our fibre mirror fabrication can be transformed to fully automated mass production, which is essential for quantum network applications such as quantum repeaters \cite{li2025, Krutyanskiy2023}.

\begin{backmatter}

\bmsection{Acknowledgments}
We thank Joel Morley and Tatsuki Hamamoto for their contributions at the early stages of the experiment, and Matthias Mader for providing us with the design for the fibre fixture. This work was supported by MEXT Quantum Leap Flagship Program (MEXT Q-LEAP) (Grant Number JPMXS0118067477) and JST CREST (Grant Number JPMJCR1873) and JST Moonshot R\&D (Grant Number JPMJMS2063). We acknowledge funding from the OIST Proof of Concept Program - Seed Phase Project (R11-59). W. J. Hughes and P. Horak acknowledge funding by the UK EPSRC Quantum Technology Hubs (EP/T001062/1 and EP/Y024389/1).

\bmsection{Data availability}
Data underlying the results presented in this paper are not publicly available at this time but may be obtained from the authors upon reasonable request.

\bmsection{The authors declare no conflicts of interest}

\bmsection{See Supplement for supporting content}

\end{backmatter}

\bibliography{fibre_shooting}

\end{document}

\maketitle

\section{Overview}

This supplement contains two main parts: the details of the cavity finesse simulations undertaken for the data presented in Figure 5 of the main text, and the procedure of preparing a copper-coated optical fibre for laser ablation.

\section{Simulation of cavity finesse}
\label{sec: simulation and fitting}
\subsection{Fitting surfaces to the profile data}
\label{sec: surface_fitting}

\begin{figure}[htbp]
\centering
\fbox{\includegraphics[width=.9\linewidth]{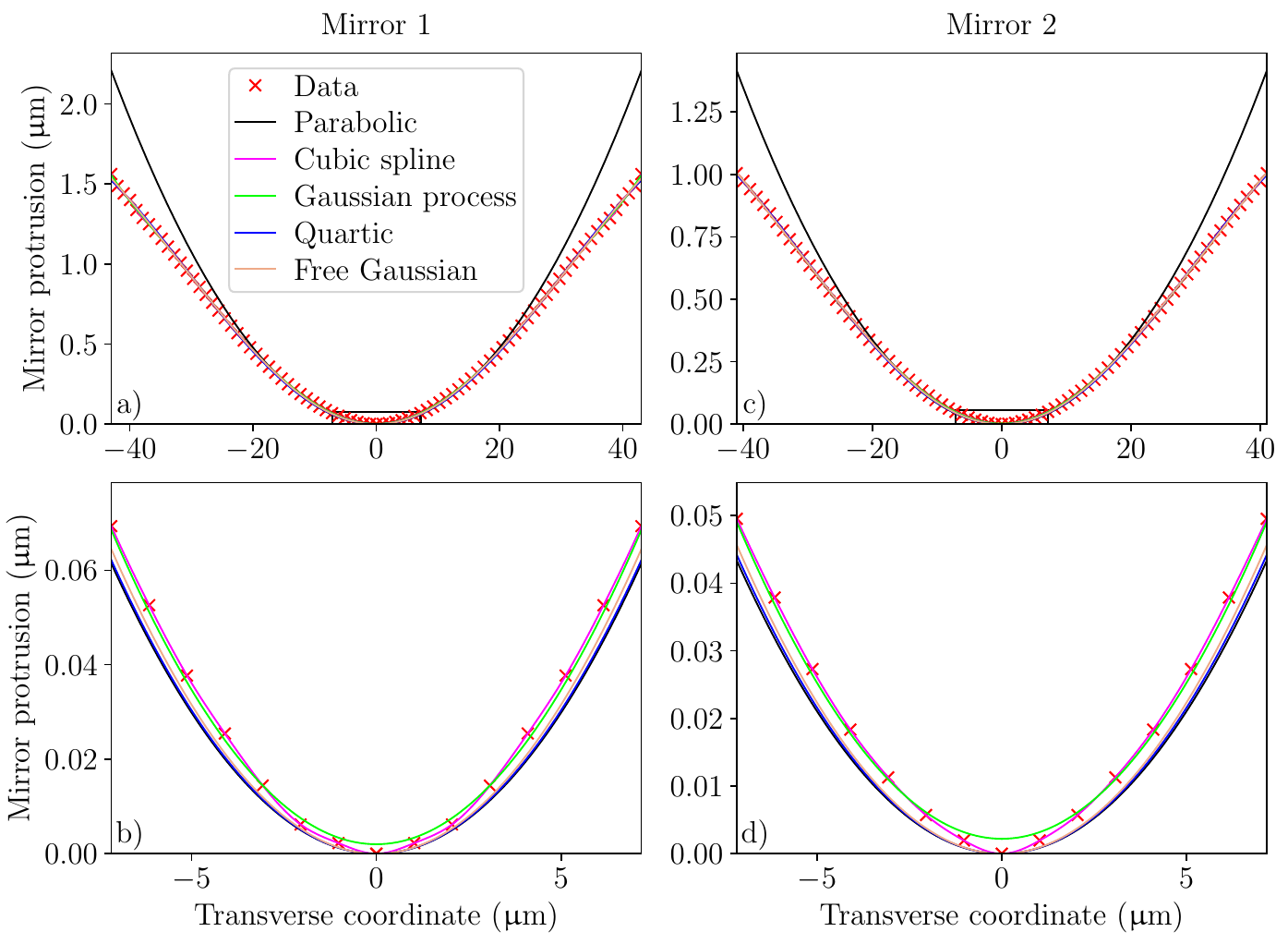}}
\caption{Different fits to the surface profiles of the mirrors that used for the cavity whose finesse is presented in Figure 5 of the main text. a) The measured protrusion as a function of radius (red crosses) with the corresponding fits (solid lines) to the profile for one mirror used in the cavity. Note that the data presented is symmetric about the origin, but is displayed for negative transverse coordinate values to emphasise the mirror shape. The rectangle in the centre indicates the zoomed region depicted b). Plots c) and d) correspond to a) and b) respectively for the other mirror of the cavity.}
\label{fig:profile_fits}
\end{figure}

The cavity whose measured finesse is presented in Figure 5 of the main text is formed of two mirrors. The mirror profiles were measured through white-light interferometry, as described in Sec. 2 of the main text, and this data was averaged over the azimuthal angle to produce mirror profiles as a function of radial coordinate from the centre. The measured radial profile, and fits to that measured data, are depicted in Figure~\ref{fig:profile_fits}. Five different fits to the profile were used:

\begin{itemize}
    \item A parabola, fitted over the central $\pm$\SI{20}{\micro\metre} of data, with the form
    \begin{equation}
        z(r) = \frac{r^2}{2R_p},
    \end{equation}
    where $z(r)$ is the fitted profile and the radius $R_p$ of the parabola is the single free parameter. A parabola is used instead of a circle for numerical simplicity in the mode calculation method discussed in Sec.~\ref{sec: simulation_cavity_finesse} of this supplement.
    \item A cubic spline, which is routed through all of the data points, and forced to have zero gradient in the centre of the profile.
    \item A fit to the surface using Gaussian process regression~\cite{Rasmussen:06}. The kernel is composed of a scaled radial basis function kernel and a white noise kernel
    \begin{equation}
        k(r_a, r_b) = A_{\mathrm{GPR}} \exp\left\{{-\frac{(r_b-r_a)^2}{l_{\mathrm{GPR}}^2}}\right\} + \mu_{\mathrm{GPR}}\delta_{ab}, 
    \end{equation}
    where $\delta$ is the Kronecker delta, $a$ and $b$ are indices that label individual data points, and the hyperparameters $A_{\mathrm{GPR}}$, $l_{\mathrm{GPR}}$, and $\mu_{\mathrm{GPR}}$ are optimised during the Gaussian process regression routine.
    \item A quartic fit to the data profile of the form
    \begin{equation}
        z(r) = \frac{r^2}{2R_q} + \alpha_qr^4,
    \end{equation}
    where $R_q$ and $\alpha_q$ are free parameters.
    \item A `Free Gaussian' fit to the profile of the form
    \begin{equation}
        z(r) = h_g\left[1-\exp\left\{-\frac{r^2}{w_g^2}\right\}\right],
    \end{equation}
    where the depth $h_g$ and waist $w_g$ are free parameters. Note that the depth $h_g$ for the fit is not constrained by the depth of the physical depression, as it is for the fits in the main text, because the intention here is purely to fit the measured data over the mirror surface. To distinguish the two conditions for the Gaussian fits, the fitting method used in this supplement is referred to as `Free Gaussian'. The central radius of curvature may also be calculated
    \begin{equation}
        R_g = \frac{w_g^2}{2h_g}.
    \end{equation}
\end{itemize}

In general terms, the parabolic fit is a decent approximation of the mirror profiles over the central region, but, with one free parameter, it does not fit the entire mirror profile well. The other fits approximate the data much better. Parameters of the fits are given in Table~\ref{tab:parameter_fits}.

\begin{table}[htbp]
\centering
\caption{\bf Fit Parameters for Mirror Surfaces}
\begin{tabular}{ccc}
\hline
Fit parameter & Mirror 1 & Mirror 2 \\
\hline
Parabolic & $R_p$=\SI{419}{\micro\metre} & $R_p$=\SI{594}{\micro\metre} \\
Quartic & $R_p$=\SI{410}{\micro\metre} & $R_p$=\SI{577}{\micro\metre}\\
 & $\alpha_q$=$-2.14\times 10^{-7}\, \upmu\mathrm{m}/\upmu\mathrm{m}^4$ & $\alpha_q$=$-1.63\times 10^{-7}\, \upmu\mathrm{m}/\upmu\mathrm{m}^4$ \\
Free Gaussian & $h_g$=\SI{2.56}{\micro\metre}, $w_g$=\SI{44.9}, ($R_g$=\SI{394}{\micro\metre}) & $h_g$=\SI{1.72}{\micro\metre}, $w_g$=\SI{43.7}{\micro\metre}, ($R_g$=\SI{556}{\micro\metre}) \\
\hline
\end{tabular}
  \label{tab:parameter_fits}
\end{table}

\begin{figure}[htbp]
\centering
\fbox{\includegraphics[width=.8\linewidth]{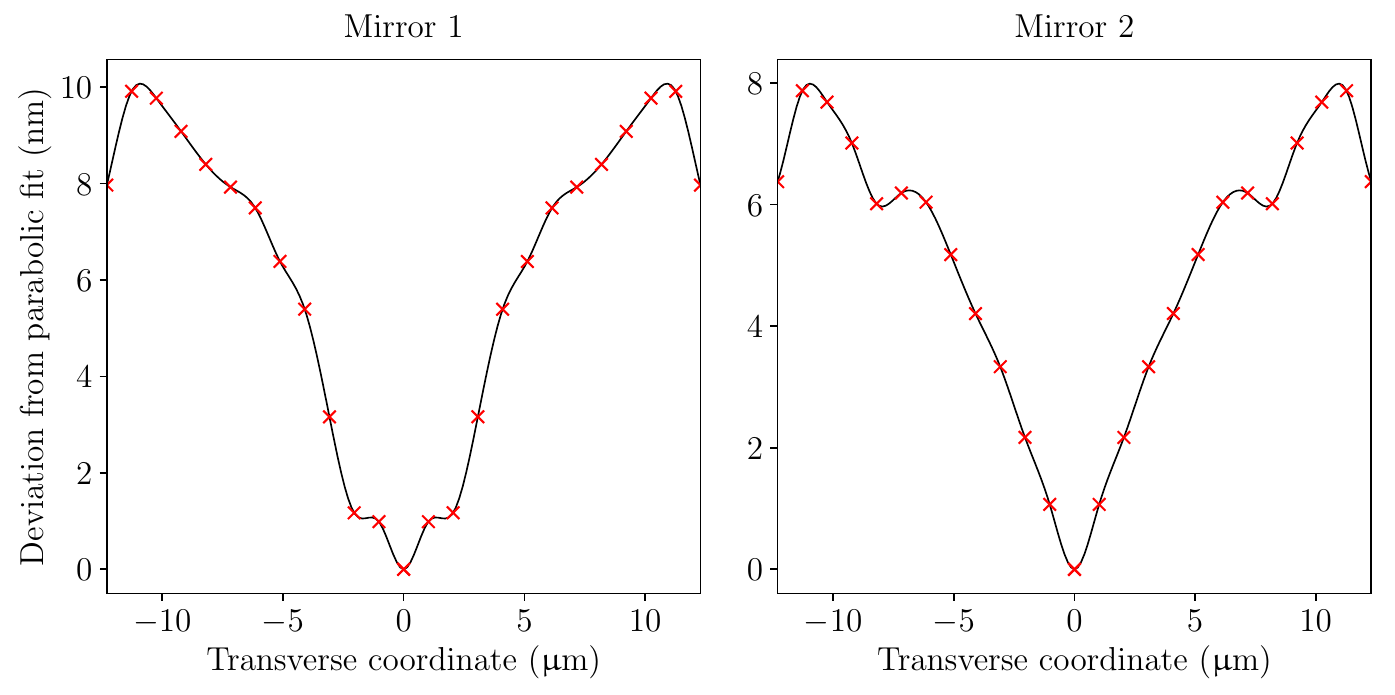}}
\caption{Spline fits (solid line) to the data (red crosses) over the centres of the mirrors. The protrusion is plotted as residuals from the parabolic fit to each mirror to highlight how the spline profile tracks noise in the measured data.}
\label{fig:spline_distortion_zoom}
\end{figure}

The cubic spline fit is unique as this surface is guaranteed to pass through all of the data points. This means that the spline profile will follow any noise in the data, as depicted in Figure~\ref{fig:spline_distortion_zoom}, which may lead to unrealistic fluctuations in the simulated surface at small length scales. The other methods are able to smooth out measurement noise through their fitting. Of these, the Gaussian process regression fit is the closest match to the data in the central region of the mirror.

\subsection{Simulating cavity finesse}
\label{sec: simulation_cavity_finesse}

The surface fits from Sec.~\ref{sec: surface_fitting} were used to simulate the expected finesse of the cavity. The cavity mode was calculated using the mode mixing method~\cite{Kleckner:10} on a Laguerre-Gauss basis containing only modes with no azimuthal dependence (due to the assumed radial symmetry of the surfaces). To include non-clipping loss (i.e. loss due to absorption in the mirror coatings, scattering from small-scale roughness, or mirror transmission), the first 3 finesse data points comprising the high-finesse plateau at low lengths were used to calculate an inverse variance-weighted average round trip non-clipping loss of $41.6\pm0.2$ppm. This non-clipping loss was divided equally between the two mirrors for the purposes of simulation (but any apportioning of losses with the correct aggregate would produce identical results).

The mode mixing method calculates a set of eigenmodes of the cavity, with none of them explicitly preferred. We pick the eigenmode with the lowest loss to be the `mode of interest' for which the finesse is calculated. This is because, firstly, this mode is likely to resemble a fundamental Gaussian beam and therefore couple well to the input and output fibre mode, and secondly, this mode has the highest number of photon round-trips, which will enhance the measured intensity. This means that the lowest loss mode is likely to be the mode with the highest intensity at the output fibre, which is the one measured in experiments (however, this is not strictly guaranteed). The finesse was calculated for the different surface fits at a range of lengths (defined as the distance between the centres of the mirrors). In each case, the mirror is assumed to have reflectivity determined by the calculated non-clipping loss within the bounds of the mirror profile and to be perfectly non-reflective outside the measured profile.

\begin{figure}[htbp]
\centering
\fbox{\includegraphics[width=.98\linewidth]{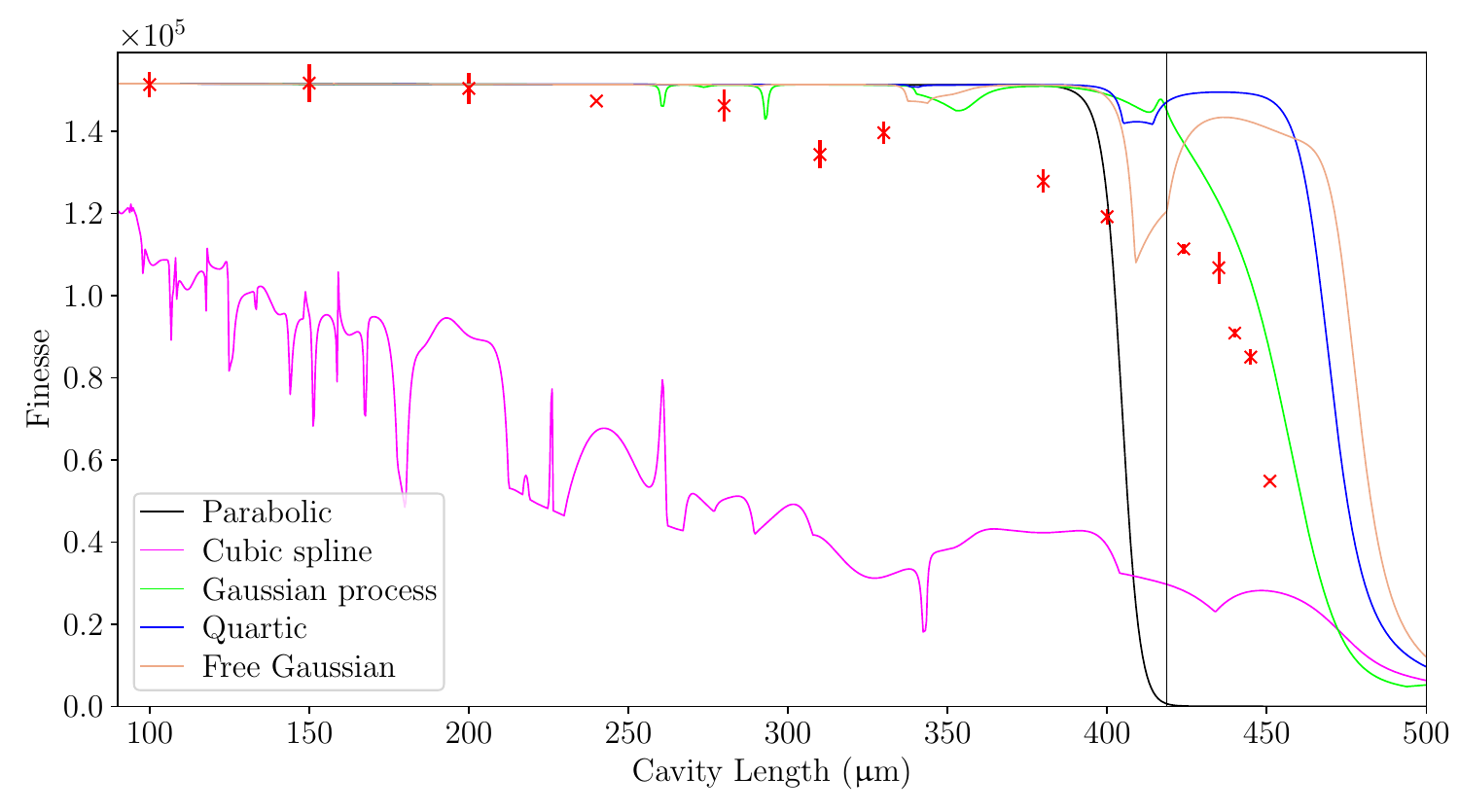}}
\caption{Measured finesse (crosses with error bars indicating the standard deviation over 6 measurements performed at different times) plotted against simulated finesse (solid lines) for different surface fitting methods as a function of cavity length. The simulations used a total of 81 basis states, with an evenly spaced 15000-sample radial grid used to calculate mode-mixing integrals. The vertical line marks the smaller radius of curvature of the parabolic fits (the larger radius of curvature is beyond the length range of the plot). Thus, at lengths to the right of the vertical line, the parabolic fit cavity is unstable.}
\label{fig:finesse_vs_length_sim}
\end{figure}

The relation between the measured data and simulated finesse for the different surface fits described in Sec.~\ref{sec: surface_fitting} is shown in Figure~\ref{fig:finesse_vs_length_sim}. This figure can be used to evaluate how appropriate different surface fitting methods are. The parabolic surface fit predicts that the cavity finesse should reduce near to zero when the length exceeds the shortest radius of curvature, but the experimental cavity still exhibited reasonable finesse at much longer lengths. The cubic spline fit features markedly reduced finesse for all bar the longest lengths. This is likely because this surface fit runs through each data point, meaning any noise on the profile measurement results in noise on the surface, as depicted in Figure~\ref{fig:spline_distortion_zoom}, which would exaggerate the true scattering loss. The other three methods (Gaussian process regression, quartic fit, and a Gaussian profile fit) agree much better with the data, but the surface generated by Gaussian process regression fits the data most closely. 

The simulated finesse does not exactly match the measured finesse. This might be because the simulated profile was constrained to have radial symmetry, whereas the true mirror surfaces will inevitably have imperfections that break radial symmetry. Another reason could be that the mirror is assumed perfectly non-reflective outside of the measured profile, whereas it may be the case that some light reflected from beyond the measured profile is retained in the cavity mode, although one would anticipate this effect to cause the simulation to underestimate the performance, in contrast with the overestimation observed.

\section{fibre fabrication procedure}
\begin{figure}[htbp]
    \centering
    \includegraphics[width=.7\linewidth]{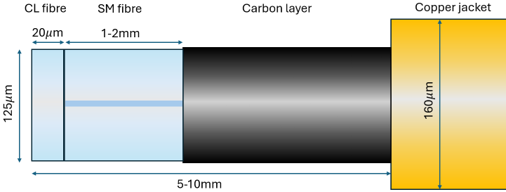}
    \caption{Schematic diagram of a fibre that has been prepared for laser ablation. 1-2 mm of carbon layer is removed of a single mode (SM) fibre, on top of which, a \SI{20}{\micro\metre} coreless fibre is attached}
    \label{fig:fibre component}
\end{figure}
For our experiments, we use Cu800SMF vacuum-compatible single-mode fibres (SMF or SM fibre) provided by IVG fibre, Toronto, Canada (www.ivgfibre.com). The fibres are primarily coated with a copper alloy and an additional inner carbon layer. The mode field diameter, cladding diameter, and coated diameter of the fibres are, respectively, \SI{6\pm 0.5}{\micro \metre}, \SI{125 \pm 1}{\micro \metre}, and \SI{165 \pm 10}{\micro \metre}. As an initial step of the bulk production of the fibre mirrors, the long fibre is cut into segments of ~1.4 meters. We prepared about a hundred fibres for the first batch of production. Figure \ref{fig:fibre component} shows a schematic diagram of Cu800 after preparation. The preparation steps are as follows.
\subsection{Removal of copper coating of the SMF}
\begin{figure}[htbp]
    \centering
    \includegraphics[width=.6\linewidth]{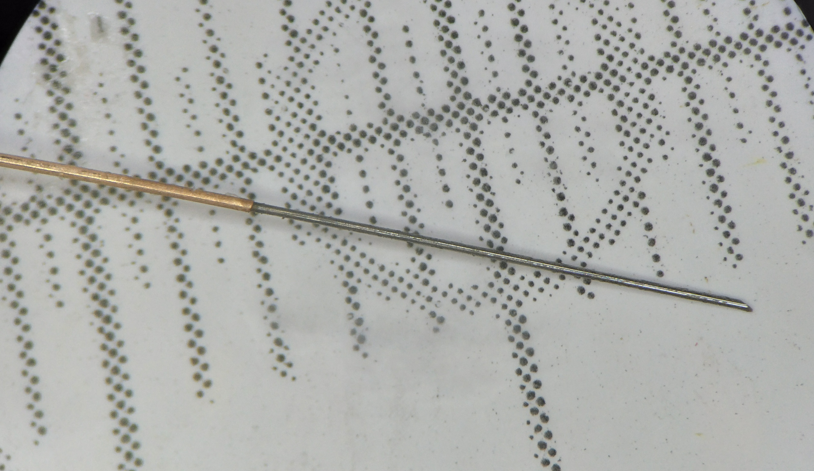}
    \caption{Image of a copper-etched SMF using FeCl3 solution.}
    \label{fig:Cu etch}
\end{figure}

To prepare the fibres for the mirror fabrication, a few centimetres of the copper coating was etched away using Iron Chloride (FeCl3) solution. The FeCl3 is dissolved in water with a one-to-four mixing ratio and heated to 50°C. Etching at room temperature also works; however, the rate of the etching process increases at the above-said temperature. A 10 cm fibre length is dipped into the solution for 15 minutes. Up to ten fibres were etched at a time because preparing more than that reduces the speed of the etching process; also, it can cause uneven etching when fibres are stacked together. Afterwards, the fibres were washed with running water and isopropyl alcohol (IPA). Figure \ref{fig:Cu etch} shows a section of Cu800SMF after some of its copper coating has been etched using the FeCl3 solution. 
\subsection{Removal of the Carbon coating of the SMF}
\begin{figure}[htbp]
    \centering
    \includegraphics[width=.6\linewidth]{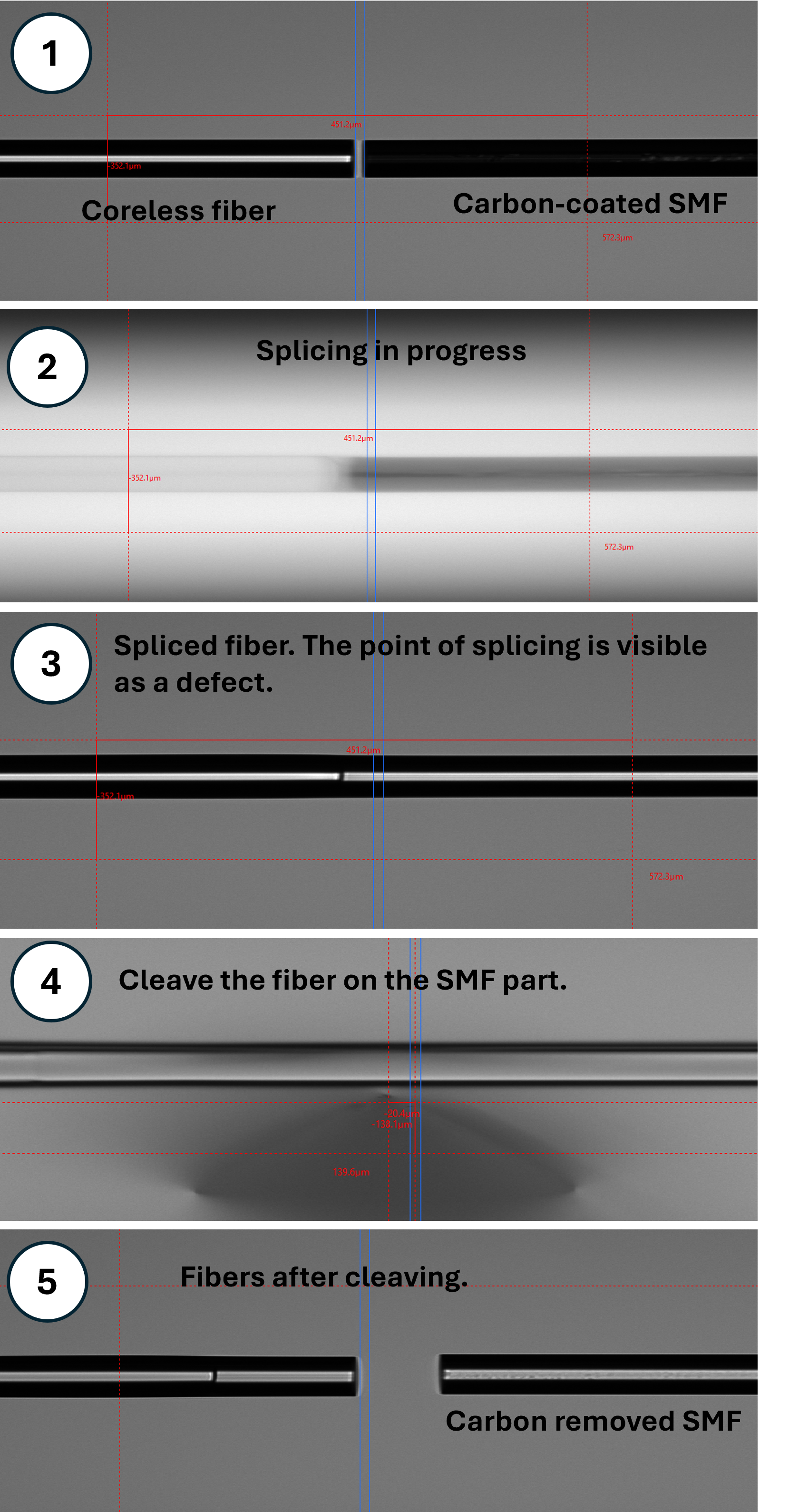}
    \caption{Steps for removing the carbon coating from the copper-etched SMF}
    \label{fig:step}
\end{figure}

The carbon layer around the SM fibre sometimes causes problems in the splicing and cleaving process, thus is recommended to be removed. There were reports about removing the carbon coating by directly heating the etched part of the fibre by applying a hot flame. However, in our experience, this method is not successful because uncontrolled flame heating causes the bending of the fibres or makes them more fragile to handle. Instead, we used the fusion splicer to remove the carbon coating. The carbon-coated fibre is spliced with a coreless fibre and then cleaved on the single-mode fibre part. The Thorlabs GPX3800 splicer, which has a built-in cleaver and a microscopic camera, provides micron-scale control over the cleaving position adjustments. During the splicing, 1-2 mm of the carbon coating of the single-mode fibre near the splicing point burns away, revealing the underneath silica fibre. After splicing, the exact splicing position is visible under the GPX3800 camera as a small discontinuity along the fibre, possibly due to carbon particles being merged with the silica. Figure \ref{fig:step} (1) to (5) show the step-by-step process of the carbon coating removal. The left part (coreless part) after cleaving is disposed of.
\subsection{Splicing of Coreless fibre to the SMF}
\begin{figure}[htbp]
    \centering
    \includegraphics[width=.6\linewidth]{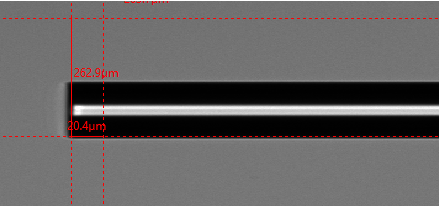}
    \caption{An image of an SMF spliced with a \SI{20}{\micro\metre} long coreless fibre}
    \label{fig:coreless}
\end{figure}

As has been explained in the main text, ablation directly on the SM fibre facet will end up with an unsmooth profile around the core. We splice a short segment of coreless fibre on top of an SM fibre to tackle this problem. A coreless fibre is spliced against the prepared single-mode fibre. With the removal of the carbon coating, defect-free splicing can be achieved at this time. The splicing intersection can be distinguished by the termination of the single-mode fibre core (the core is visible under the camera as a thin dark line in the middle of the fibre). Then, with the help of the in-built cleaver and the microscope camera, a \SI{20}{\micro\metre} length of the coreless fibre is cleaved from the splicing point. So we have \SI{20}{\micro\metre} length of the coreless fibre on top of the SM fibre. This is the final step of preparing the fibres for laser ablation. Figure \ref{fig:coreless} shows an SMF spliced with \SI{20}{\micro\metre} of coreless fibre.
\subsection{Loading the prepared fibres to the fibre fixture}

\begin{figure}[htbp]
    \centering
    \includegraphics[width=1\linewidth]{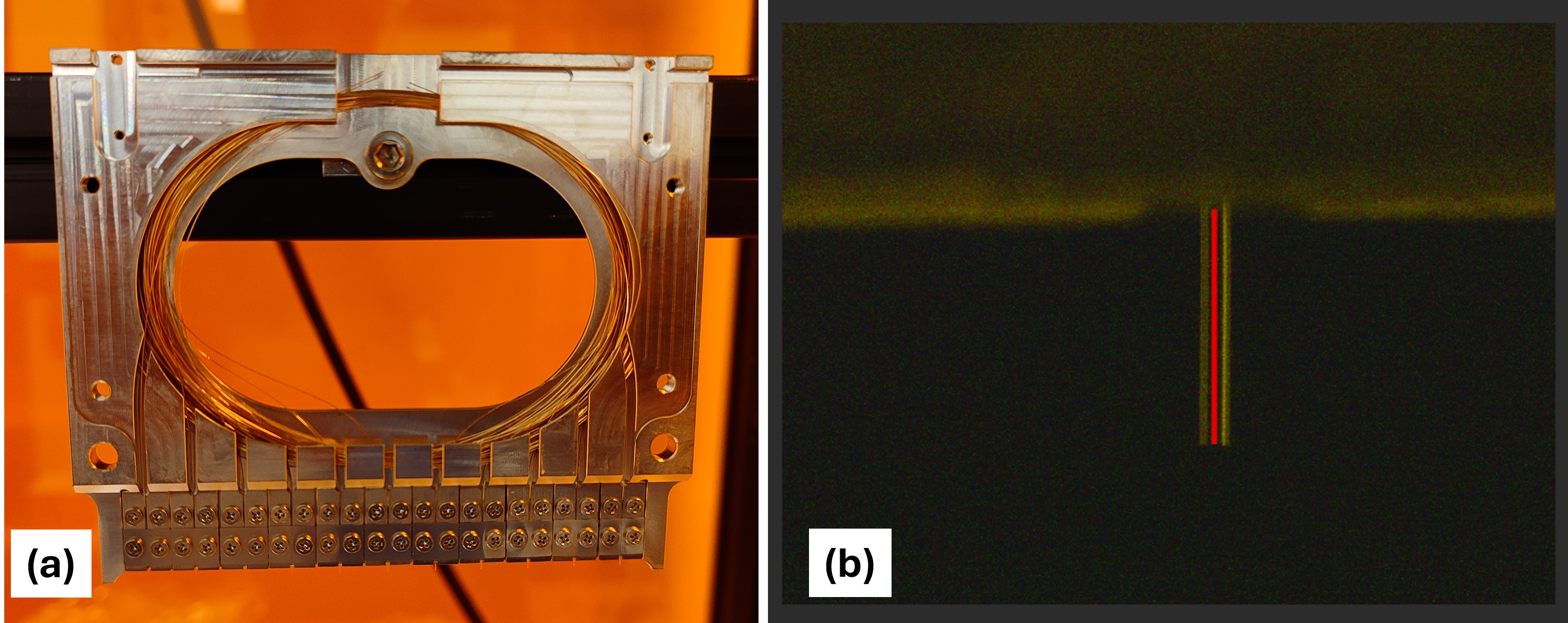}
    \caption{(a)Image of a fibre-loaded fixture (b) Close image of the fibre protruding out of the holder. The protrusion is 1mm}
    \label{fig:fibre loading}
\end{figure}

After preparing the fibres, we individually store them on a homemade fibre fixture designed initially by Dr. Matthias Mader, Ludwig-Maximilians-University, Munich, Max-Planck-Institute of Quantum Optics Garching. The fibre fixtures were designed to allow precise positioning of the fibre end facets to achieve uniformity coating in the Ion Beam Sputtering (IBS). The original form of the fibre fixture is modified according to our requirements. The loading process is undertaken inside a laminar flow box to prevent dust deposition on the fibre end facets. One fibre fixture can hold eleven fibres, and an assembly of five fixtures forms a block, which can be stably mounted in the IBS coating chamber. Before loading the fibres, all the fibre fixtures, including the fibre grooves, fibre clamps, screws etc., are cleaned in the ultra-sonic bath, first with soap solution and then with acetone and Isopropyl alcohol (IPA) for 15 minutes. It is then dried using nitrogen airflow. Fibre fixing was done under a microscope, which helps to precisely adjust so that the fibre protrudes by 1~mm from the edge of the fibre fixture. All fibres should protrude the same length with a precision of less than \SI{100}{\micro\metre}, which guarantees high-reflective coating for the correct wavelength (850 nm). The remaining lengths of the fibres were wound up and stored in the race-track-like storage slit within the fibre fixture. Figure \ref{fig:fibre loading} shows an image of a fibre fixture with loaded fibres.
\subsection{Loading the fibre fixtures to the laser ablation unit}

\begin{figure}[htbp]
    \centering
    \includegraphics[width=0.6\linewidth]{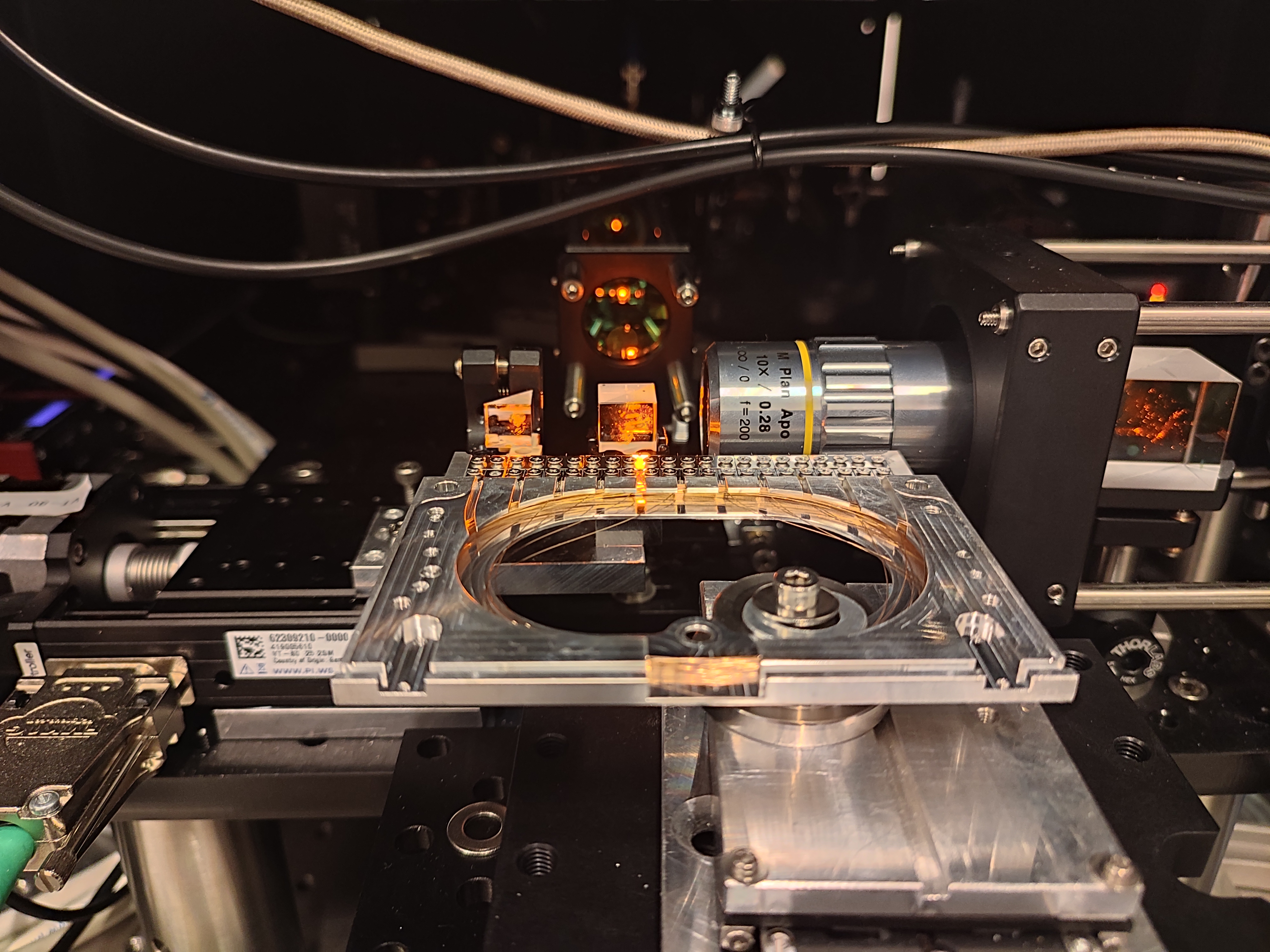}
    \caption{Image of the fibre fixture loaded on the laser ablation unit}
    \label{fig:fixture loading}
\end{figure}

The prepared fibre-loaded fixture is installed on the laser ablation unit. Before installing, the fibre-loaded fixture is carefully washed in an ultrasonic bath, first with acetone and then with IPA. Each wash lasts 5 minutes. This is to ensure the cleanness of the fibres just before the ablation.  Figure \ref{fig:fixture loading} shows the image of a fibre fixture loaded on the laser ablation unit. After that, laser shooting of the fibres takes place one by one. Eleven fibres in one fixture can be laser machined all at once by successively translating the fixture horizontally with the linear stage. The surface profile data for each fibre are labelled for future reference. After ablating all the fibres, the fixture was moved to a packaging assembly prepared inside a vacuum desiccator to minimize contamination.  
\subsection{Packaging of the laser-machined fibres for the IBS coating}
\begin{figure}[htbp]
    \centering
    \includegraphics[width=0.75\linewidth]{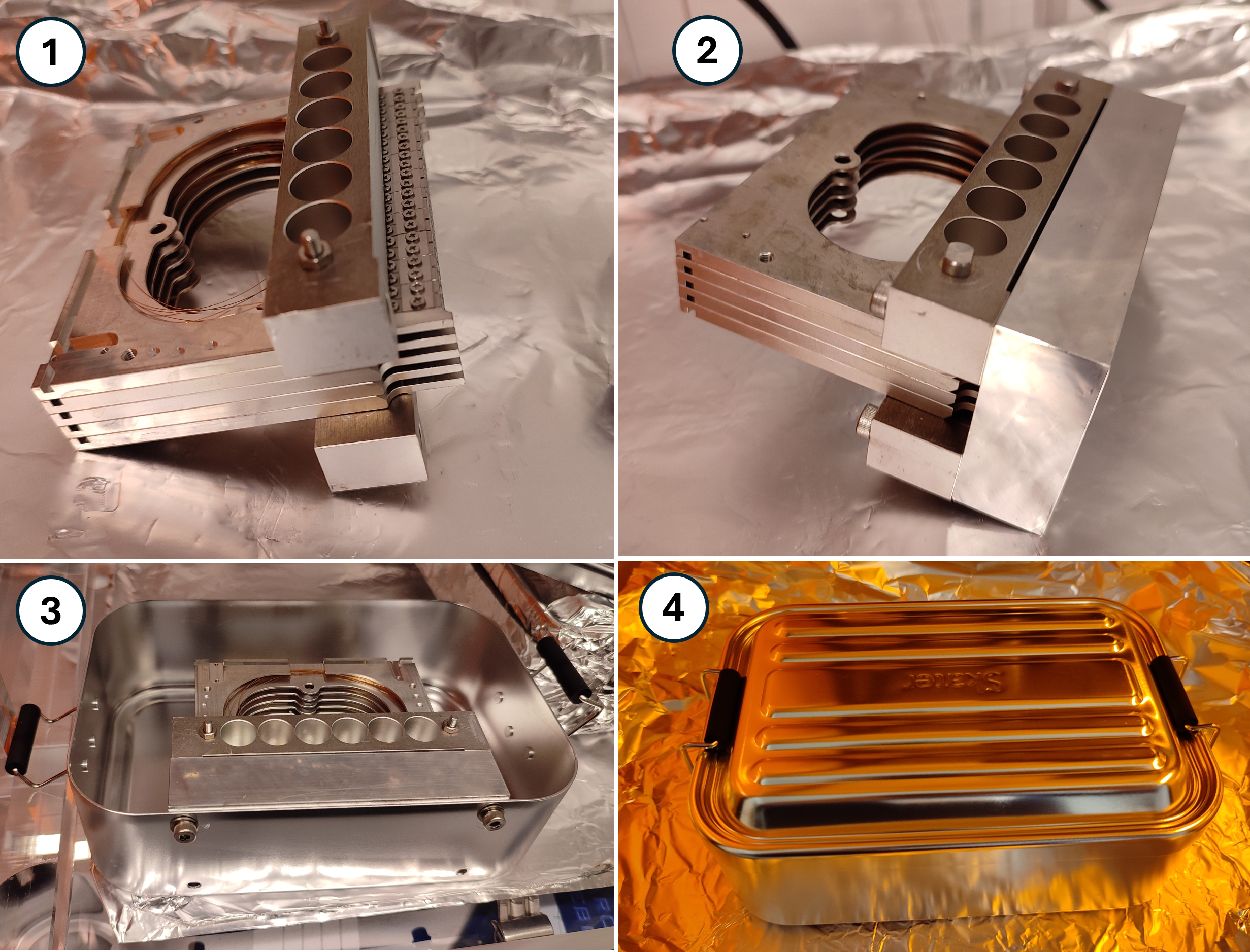}
    \caption{(1) A fibre block for IBS coating (2) Facets of the fibres are protected with a front cover (3) The secured block is installed inside a cleaned metal box (4) A packaged block for transportation.}
    \label{fig:for shipping}
\end{figure}

After laser machining the fibres, the fibre fixtures were assembled as a block, which helps to securely handle the fibres for transportation and install them in the IBS coating chamber. The fibre fixture blocks were securely packed inside a cleaned metal container, and then aluminium foil, before being transferred to the company, LASEROPTIK GmbH, a provider of high-reflective IBS coating.  All the assembling took place inside a flow box to reduce the chances of any contamination.  Figure \ref{fig:for shipping} is an image of a fibre fixture block after the assembly.

\bibliography{fibre_shooting_suppl}